\documentclass[10pt,aps,prx,twocolumn,amsmath,amssymb,superscriptaddress,nobibnotes]{revtex4-2}

\usepackage{booktabs} 
\usepackage{siunitx} 

\usepackage{graphicx}
\usepackage{dcolumn}
\usepackage{bm}

\usepackage[hidelinks]{hyperref}
\usepackage{verbatim}
\usepackage{multirow}
\newcommand{\angstrom}{\mbox{\normalfont\AA}}
\begin{document}

\title{Microscopic Theory of Density Scaling: Coarse-Graining in Space and Time}
\author{Jaehyeok Jin}
\email{jj3296@columbia.edu}
\affiliation{Department of Chemistry, Columbia University, New York, New York 10027, United States}
\author{David R. Reichman}
\email{drr2103@columbia.edu}
\affiliation{Department of Chemistry, Columbia University, New York, New York 10027, United States}
\author{Jeppe C. Dyre}
\email{dyre@ruc.dk}
\affiliation{``Glass and Time'', IMFUFA, Department of Science and Environment, Roskilde University, P.O. Box 260, DK-4000 Roskilde, Denmark}
\author{Ulf R. Pedersen}
\email{ulf@urp.dk}
\affiliation{``Glass and Time'', IMFUFA, Department of Science and Environment, Roskilde University, P.O. Box 260, DK-4000 Roskilde, Denmark}

\date{\today}

 \begin{abstract} Understanding the structure and dynamics of liquids is pivotal for the study of larger spatiotemporal processes, especially in glass-forming materials at low temperatures. Density scaling, observed in many molecular systems through experiments, offers an efficient means for exploring a vast range of time scales along a one-dimensional phase diagram. However, the theoretical foundation provided by isomorph theory is of limited use for molecular systems, since currently no first-principles theory exists that can explain the origins of density scaling or make predictions based on it. In this work, we propose a first-principles framework employing coarse-graining in space and time. Spatial coarse-graining reduces a molecule to a center-of-mass-level description by eliminating fast degrees of freedom, while temporal coarse-graining involves averaging fluctuations or correlation functions over characteristic time scales. We show that both approaches enable \textit{ab initio} estimation of the density scaling coefficient for ortho-terphenyl, consistent with experimental values. Building on these findings, we employ excess entropy scaling to derive a microscopic theory that underpins density scaling from fully atomistic simulations. Our results illuminate the role of coarse-graining in assessing slow fluctuations in molecules and unravel the microscopic nature of density scaling. Ultimately, our proposed framework enables systematic bottom-up approaches for predicting transport coefficients that are otherwise experimentally inaccessible and computationally prohibitive. 
\end{abstract}

\maketitle

\section{Introduction} \label{sec:Introduction}
First-year courses in thermodynamics and statistical physics teach the important fact that the phase diagram of pure substances is two-dimensional \cite{Chandler1987, Schroeder2000}. However, experimental findings at elevated pressures have revealed that the thermodynamic phase diagram can sometimes be reduced to one dimension \cite{Roland2005, Niss2018}. 
The density scaling of dynamics, in particular, has been observed in experiments involving glass-forming liquids and various other fluids \cite{Tolle1998, Tolle2001, Dreyfus2003, AlbaSimionesco2002, Roland2005, Fragiadakis2011, Pawlus2020, Niss2018, Hansen2018b, Hansen2020, Koperwas2020} and follows the relationship 
\begin{equation}\label{eq:densityscaling}
\rho^\gamma/T=\mathrm{(constant)},
\end{equation}
where the constant is related to transport properties, for instance, the diffusion coefficient $D$ in reduced units, $\rho=N/V$ is the number density, $T$ is the temperature, and $\gamma$ is the \textit{density scaling exponent}. In other words, Eq. (\ref{eq:densityscaling}) implies that the ratio $\rho^\gamma/T$ remains invariant for different values of $T$ and $\rho$ as long as $D$ is the same.

Remarkably, Eq. (\ref{eq:densityscaling}) links a difficult-to-predict dynamical property, $D$, to thermodynamic properties via the quantity, $\rho^\gamma/T$, and this connection thus opens up avenues for both practical and theoretical exploration. In practical terms, density scaling offers a more efficient means of computing phase diagrams by reducing dimensionality. It also provides estimates of transport coefficients in regions where experiments are difficult, or simulations are computationally expensive. This is particularly valuable for glass-forming liquids near the glass transition, where properties change over several orders of magnitude \cite{kauzmann1948nature, ediger1996supercooled, angell2000relaxation} and could potentially be extended to the long-wavelength dynamics of biomembranes \cite{pedersen2010correlated}. Equally important, density scaling enables the exploration of deeper theoretical implications of hidden scale invariance. Notably, it offers insights into the microscopic origins of the dynamical slowdown in glass formers, from the perspective of growing point-to-set structural correlations as a liquid approaches the glass transition \cite{hocky2012growing}. By addressing these fundamental questions, density scaling not only provides a practical framework for predicting the behavior of liquids, but also opens new avenues for theoretical exploration and understanding.  

Despite its experimental prevalence, understanding density scaling from first principles remains a major challenge. This limitation greatly hinders the applicability of scaling for complex systems. This is due to two major issues. First, density scaling was initially discovered through experimental observations, and there is currently no microscopic theory capable of explaining Eq. (\ref{eq:densityscaling}) or deriving the scaling exponent $\gamma$ from first principles. 
Existing work largely focuses on toy models, which are only applicable to a limited set of liquids, e.g., for systems with inverse power-law (IPL) interactions, $v(r)=\varepsilon (r/\sigma)^{-m}$, where $\gamma=m/3$ conforms exactly to density scaling  \cite{Hoover1971, Hoover1972}. However, these simple models are not representative of more realistic and experimentally relevant molecules, such as most glass-forming liquids and biological systems. Moreover, even when scaling does hold, Eq. (\ref{eq:densityscaling}) does not specify the exact form of the scaling relationship. Specifically, while Eq. (\ref{eq:densityscaling}) indicates that $\rho^\gamma/T$ remains a function of the self-diffusion coefficient, $D$, it does not provide an explicit relationship between $D$ and $\rho^\gamma/T$. To date, no attempt has been made to derive this explicit relationship from first principles.

An analytical link between transport properties and thermodynamic properties is established through another scaling relationship known as \textit{excess entropy scaling} in liquids. Initially proposed by Rosenfeld in 1977 \cite{Rosenfeld1977}, excess entropy scaling relates dimensionless transport properties, denoted as $\tilde{D}$, to the per particle excess entropy of the system, $s_\mathrm{ex}:=S_\mathrm{ex}/Nk_B$,
\begin{equation}\label{eq:excessscaling}
\tilde{D} = D_0 \exp (\alpha s_\mathrm{ex}),
\end{equation}
where \emph{excess} refers to the entropy in excess of the ideal gas entropy at the same density $\rho$ and temperature $T$ \cite{theory_of_simple_liquids, Dyre2018}: $S_\textrm{ex}(\rho, T)=S(\rho, T)-S_\textrm{id}(\rho, T)$. Equation (\ref{eq:excessscaling}) has been shown to hold for various simple analytical systems such as Lennard-Jones \cite{Jones1924}, IPL \cite{Hoover1971, Hoover1972}, and hard spheres \cite{Bernal1960}, as well as a wide range of molecular systems \cite{Li2005, Errington2006,krekelberg2007,chakraborty2007,Krekelberg2009,Chopra2010,Mausbach2018,Hopp2018,Dyre2018,Costigliola2018,Costigliola2019,Bell2019,Bell2020b,Harris2020, Douglas2021, LotgeringLin2015, LotgeringLin2018, Baled2018, Rokni2019, BintiMohdTaib2020, Liu2020, Bell2020_I, Bell2020_II, Yang2021, Bell2022}. Excess entropy scaling can even be extended beyond the atomistic level to coarse-grained (CG) models \cite{jin2023understanding,jin2023understanding2,jin2023understanding3, 10.1063/5.0212973} and out-of-equilibrium systems \cite{ghaffarizadeh2022excess, saw2023configurational}.

As excess entropy depends on both density and temperature, excess entropy scaling implies that dimensionless transport properties may be directly related to these two thermodynamic variables. Therefore, we posit that excess entropy scaling provides a viable route to link $\tilde{D}$ and $\rho^\gamma/T$. Furthermore, despite the empirical nature of scaling, recent work has shown that Eq. (\ref{eq:excessscaling}) can be derived from hard sphere theory \cite{widom1967intermolecular} by coarse-graining molecules to effective hard spheres \cite{jin2023understanding,jin2023understanding2,jin2023understanding3, 10.1063/5.0212973}, indicating that such a link can, in principle, be analytically derived from microscopic physics \cite{mirigian2013unified,mirigian2014elastically1,mirigian2014elastically2,mirigian2015dynamical,mei2021experimental1,mei2021experimental2}.

In this paper, we aim to unravel the microscopic origins of density scaling from first principles. First, we extend the range of isomorph theory, which provides a physical framework for understanding the scaling exponent but has been mostly successful for simple atomic models \cite{scl_IV,scl_V}, to more realistic molecular liquids by introducing the concept of systematic coarse-graining. By constructing a CG representation of complex molecules across both spatial and temporal dimensions, we greatly broaden the applicability of isomorph theory to realistic molecular liquids to gain a microscopic understanding of density scaling exponents. Next, we will integrate excess entropy scaling with density scaling to analytically derive a microscopic theory underlying density scaling. Accurately determining excess entropy for complex molecules is presently challenging, but we will address using the developed CG framework.

Ortho-terphenyl (OTP), a quintessential glass-forming liquid, serves as an ideal candidate for this line of research, as it has been shown to exhibit density scaling in experiments \cite{Greet1967, Richert2005, Takahara1999, Casalini2016, Mapes2006}. However, the experimental scaling coefficients observed differ significantly from those predicted by simple model potentials, i.e., Lennard-Jones potential ($m=12$ with $\gamma=4$). Currently, no first-principles approach exists to derive this exponent. We thus focus on this paradigmatic case, although our approach applies generally to complex systems.

The remainder of this paper will demonstrate that our approach achieves three key objectives. First, it allows one to determine whether density scaling is valid for a given system, which is important for understanding and predicting hidden scale invariance in complex molecules. Second, it enables one to estimate scaling exponents \textit{ab initio}, facilitating the efficient construction of phase diagrams. Finally, by leveraging excess entropy scaling from a microscopic perspective, our approach makes it possible to predict transport coefficients in regimes that are difficult to simulate, such as supercooled liquids near the glass transition. These achievements address significant challenges that have historically created a divide between experimental observations and theoretical predictions. This paper ultimately aims to bridge this gap.

\section{Isomorph Theory}
Isomorph theory is a theoretical framework that explains density scaling in a class of liquids, the so-called ``strongly correlating'' liquids \cite{scl_IV,scl_V}. In this section, we provide a brief overview of isomorph theory and delineate its current limitations, particularly in the context of molecular systems. Isomorph theory primarily centers on identifying invariant structures and dynamics within atomic systems, and hence we start from the total energy as a summation of kinetic and potential energies for a system of $n$ particles: 
\begin{equation}
    E(\dot {\bf r}^n, {\bf r}^n) = K(\dot {\bf r}^n)+U({\bf r}^n),
\end{equation}
where $ {\bf r}^n \equiv ( {\bf r}_1, {\bf r}_2, {\bf r}_3,\ldots,{\bf r}_n)$ is the $3n$-dimensional fine-grained configurational variables for atomic positions. The fundamental assumption of isomorph theory is that the potential energy function $U({\bf r}^n)$ is scale-invariant in the following generalized sense \cite{Mandelbrot1968, Guth1982, Hawking1982, Mandelbrot1983, Pelissetto2002, Bak2002, Schroeder2014}.
Consider two configurations with the same density, ${\mathbf r}_a^n$ and ${\mathbf r}_b^n$, where
$
	U({\bf r}_a^n)<U({\bf r}_b^n).
$
If the energy surface is scale-invariant for these configurations, it follows that 
$
	U(\lambda {\bf r}_a^n)<U(\lambda  {\bf r}_b^n),
$
where $\lambda$ determines the magnitude of an affine scaling of all particle positions. This property can be envisioned as the energy surface \emph{retaining its shape} when the length scale changes \cite{Dyre2014}. 

Scale invariance is trivial if the energy surface, $U({\bf r}^n)$, is Euler homogeneous, for example a sum of IPL pair interactions ($\propto r^{-m}$) with the same exponent $m$ \cite{Hoover1971,Hoover1972, Simionesco2004,Casalini2004,Roland2005,Pedersen2008, Pedersen2010b}. Unlike a single-exponent interaction, e.g., Coulombic ($r^{-1}$) as in plasma, the Lennard-Jones model is a sum of two power laws, and hence the scaling is not trivial and inexact, but merely a good approximation, which implies that one can obtain an approximate scaling exponent that varies with state points. This property is referred to as \emph{hidden scale invariance}.

In systems with hidden scale invariance, there are lines in the phase diagram along which structure, dynamics, and certain thermodynamics quantities are invariant to a good approximation when given in the so-called reduced units. These units are state point-dependent and defined as a combination of particle mass $m$, number density $\rho$, and thermal energy $k_BT$ \cite{scl_IV}. Thus, the length unit is the average interparticle distance, the energy unit is the thermal energy, and the time unit is the time it takes to move an interparticle distance with thermal velocity. The lines of invariance, known as ``isomorphs,'' are configurational adiabats, i.e., characterized by $S_\textrm{ex}=\mathrm{(constant)}$. It is this property that connects isomorph theory to excess entropy scaling.

Hidden scale invariance can be validated by quantifying the correlations between the fluctuations of the virial $W$ and the potential energy $U$. In the canonical ensemble \cite{Heyes1998, Frenkel2002}, the correlation coefficient $R$ is given by \cite{Pedersen2008} 
\begin{equation}\label{Eq:R}
	R(\rho, T) =  \frac{\langle\Delta W\Delta U\rangle}{\sqrt{\langle(\Delta W)^2\rangle\langle(\Delta U)^2\rangle}},
\end{equation}
where $\Delta$ refers to deviations from the mean value. Here, the virial $W$ is the \emph{configurational contribution} to the pressure ($p$), defined via $pV=Nk_BT+W$, and $W=\langle W({\bf r}^n)\rangle$ is the \textit{NVT} (canonical) ensemble average of the instantaneous virial. This quantity can be computed for a given configuration ${\bf r}^n$ as the change in energy of an affine scaling,
\begin{equation}
    W({\bf r}^n) = \left( \frac{\partial U({\bf r}^n)}{\partial \ln \rho} \right)_{\bf{\tilde r}^n},
\end{equation}
where $\tilde {\bf r}^n = {\bf r}^n\rho^\frac{1}{3}$ is the reduced coordinates (keeping $\tilde {\bf r}^n$ constant during a density change amounts to performing a uniform scaling of all coordinates). From Eq. (\ref{Eq:R}), if $W({\bf r}^n)$ and $U({\bf r}^n)$ are perfectly correlated, then $R=1$. 

For a perfectly scale-invariant interaction $U({\bf r}^n)$, one has $R=1$. Systems with $R$ close to unity exhibit hidden scale invariance. The pragmatic criterion defining ``strong correlation'' is $R>0.9$ \cite{scl_I}. Generally, $R$ is a function of both density and temperature: $R(\rho, T)$, and hidden scale invariance applies only in some parts of the phase diagram. For Lennard-Jones systems, for example, it is generally found that $R(\rho, T)$ is low near the gas-liquid coexistence as well as in the gas phase, while it is close to unity in the dense liquid and crystalline phases \cite{scl_I}.

For atomic systems, isomorphs are identified as lines where the excess entropy $S_\textrm{ex}$ is constant, representing a configurational adiabat. The ``slope'' of the configurational adiabat in the logarithmic density-temperature phase diagram is 
\begin{equation}\label{eq:gamma}
	\gamma(\rho, T) := \left.\frac{\partial \ln T}{\partial \ln \rho}\right|_{S_\textrm{ex}}\,.
\end{equation}
A generally valid thermodynamic identity implies that this can be evaluated in a canonical simulation as \cite{scl_IV}
\begin{equation}\label{eq:gamma2}
	\gamma(\rho, T) = \frac{\langle\Delta W\Delta U\rangle}{\langle(\Delta U)^2\rangle}.
\end{equation}
Whenever density scaling holds as a good approximation, Eq. (\ref{eq:gamma2}) provides a method for determining the exponent $\gamma(\rho, T)$ from simulations \cite{Gundermann2011}. Note, however, Eq. (\ref{eq:gamma2}) may exhibit a significant state point dependence, in which case the density scaling approximation will fail to yield a constant exponent.

As mentioned, the isomorph theory is exact only when the potential energy function $U({\bf r}^n)$ is Euler-homogeneous. In the case of the IPL ($\propto r^{-m}$) potential, virial and potential energy fluctuations are perfectly correlated, expressed as
\begin{eqnarray}
    \Delta W(t)=\gamma^\ast \Delta U(t),
\end{eqnarray}
with $\gamma^\ast = m/3$ and a density scaling of the form $\rho^{\gamma^\ast}/T$. 

\begin{figure}
\includegraphics[width=0.7\columnwidth]{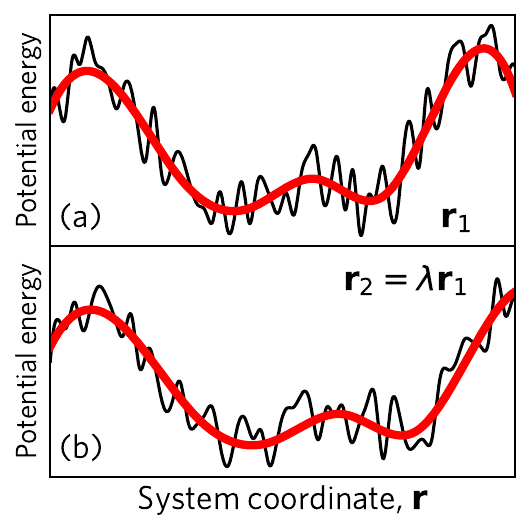}
\caption{\label{fig:energy_landscape} Schematic illustration of our coarse-graining approach for understanding hidden scale invariance. (a) A conceptual sketch of the typical potential energy landscape, $U({\bf r}^n)$, for a molecular system. The landscape exhibits two characteristic length scales: (1) fast fluctuations (black curve) dominated by intramolecular vibrations and intermolecular roughness and (2) slow fluctuations (red curve) representing substantial molecular rotations and translations. (b) The potential energy landscape after an affine scaling of all coordinates by 10\% ($\lambda=1.1$). Due to fast fluctuations, the potential energy landscape is not intrinsically scale-invariant. However, the slow fluctuations are scale-invariant, i.e., the red curve retains its shape. This paper addresses the challenge of identifying the solid red line from the black line in an atomistic description of a molecular system.}
\end{figure}

In its most used form, the isomorph theory is formulated for atomic systems (i.e., point particles), but it often breaks down when applied to molecular systems \cite{Pedersen2010c, Ingebrigtsen2012, Dyre2014, Veldhorst2015, Koperwas2020b}. This breakdown is often due to the existence of strong directional intramolecular interactions, e.g., covalent or hydrogen bonds destroying the strong virial-potential correlations, resulting in low $R$ values [see Fig.\ \ref{fig:energy_landscape}(a) and \ref{fig:energy_landscape}(b)].
The effect of intramolecular vibrations is another factor that generally significantly weakens the correlations between virial and potential energy. This becomes evident when, compared to a rigid Lennard-Jones chain \cite{Veldhorst2014}, the flexible chain described by harmonic bonds manifests the breakdown of the isomorph theory with an instantaneous virial-energy correlation coefficient of only 0.28 \cite{Veldhorst2015}. Thus, the potential energy surface of this flexible molecule is not scale invariant, and one could only understand isomorph theory for molecules in an \textit{ad hoc} manner. For example, isomorph theory cannot be directly applied to alkane molecules; instead, one could approach it using an \textit{ad hoc} CG Lennard-Jones chain description. As the existing literature lacks studies that rigorously bridge microscopic interactions and isomorph theory at the \textit{molecular level}, the main objective of this paper is to explore how we can apply hidden scale invariance to molecules, thereby expanding the range of isomorph theory. 

\section{Systematic Coarse-Graining Approaches}
\subsection{Need for Bottom-up Coarse-Graining}
Due to the strong intramolecular interactions, the instantaneous virial and potential energy are expected to exhibit significant fluctuations at the atomistic level. In order to overcome this challenge, we propose statistical coarse-graining approaches that allow for extracting an effective one-dimensional phase diagram for these complex molecules. Our central hypothesis is that the correct coarse-graining procedure, derived from microscopic physical principles, can effectively integrate out unnecessary degrees of freedom, e.g., intramolecular vibrations, and retain only the important intermolecular motions, which in many cases will have strong correlations and, thus, resulting in an effectively one-dimensional thermodynamic phase diagram. 

In this section, we introduce two separate coarse-graining methodologies: coarse-graining in time and in space. Since these two approaches are derived from different underlying assumptions, we develop each scheme separately, apply them to the atomistic OTP, and compare their performances in Secs. \ref{sec:Coarse-graining-in-time} and \ref{sec:Coarse-graining-in-space}. Then, combining these two coarse-graining approaches in Sec. \ref{sec:Outlook}, we assess how systematic coarse-graining approaches impart the one-dimensional phase diagram via density scaling.

\subsection{Coarse-graining in time}

Coarse-graining in time (or \textit{temporal coarse-graining}) mitigates the fluctuations in the fully atomistic potential energy landscape by averaging the fluctuations within a characteristic time scale. This approach enables the examination of slow degrees of freedom, which can be measured, e.g., by frequency-dependent linear response functions. 

The central idea underlying temporal coarse-graining is motivated by an experiment on the silicone diffusion pump oil DC704 (tetramethyltetraphenyl-trisiloxane; C$_{28}$H$_{32}$O$_2$Si$_3$) \cite{Gundermann2011}. In contrast to atomistic simulations, it is impractical to measure the correlation coefficient $R$ defined by the instantaneous virial-potential fluctuations from experiments. Instead, Ref. \onlinecite{Gundermann2011} utilized the fluctuation-dissipation theorem \cite{theory_of_simple_liquids}, providing an alternative relationship to dynamic response functions, e.g., the frequency-dependent heat capacity. Extending the isomorph theory to frequency-dependent response functions enables scaling to be applied to the slow degrees of freedom, i.e., molecular rotational and translational motions, which freeze at the glass transition \cite{Goldstein1969, Ediger2012, Biroli2013}. This approach generalizes the static correlation coefficient of the isomorph theory for point particles, $R$ [Eq. (\ref{Eq:R})], to a time-dependent equivalent, $R(t)$, that is accessible by experiments. In detail, experimental measurements have demonstrated that, to the first approximation, the long-time limit of $R(t)$ is related to the classic Prigogine-Defay ratio $\Pi$,
a dimensionless number, defined as
\begin{eqnarray} \label{eq:prigogine_classic}
    \Pi \equiv \frac{1}{VT}\frac{\Delta c_p \Delta \kappa_T}{(\Delta \alpha_p)^2}\Biggr|_{T=T_g} \simeq \frac{1}{R^2},
\end{eqnarray}
which involves the non-trivial jumps ($\Delta$) in the thermal expansion coefficient $\alpha_p$, compressibility $\kappa_T$, and heat capacity $c_p$ at the glass transition $T=T_g$  \cite{chemical_thermodynamics, Gupta1976, Takahara1999, Schmelzer2006, Ellegaard2007, Pedersen2008b, Gundermann2011, Fragiadakis2011, Casalini2011, Tropin2012, Garden2012}, i.e., $\Delta c_p := \lim_{T\rightarrow T_g+} c_p - \lim_{T\rightarrow T_g-}c_p$, and similarly for $\Delta \kappa_T$ and $\Delta\alpha_p$. $\Pi$ is unity whenever the phase diagram is effectively one-dimensional.

By establishing $\Pi \simeq {1}/{R^2}$ [Eq. (\ref{eq:prigogine})] for DC704, Ref. \onlinecite{Gundermann2011} further sheds light on the strongly correlating nature of OTP in experiments. Estimating the classical Prigogine-Defay ratio for 22 glass formers from literature values, including polymers, metallic alloys, inorganic, and molecular liquids (both hydrogen-bond rich and van der Waals bonded), revealed that liquid mixtures containing OTP have $\Pi$ values near 1.2, corresponding to $R\simeq0.9$. This indicates the existence of an effective one-dimensional phase diagram of OTP. Notably, this back-of-the-envelope estimation correctly predicts the behavior of hydrogen-bonding liquids, where glucose (C$_6$H$_{12}$O$_6$) and glycerol (C$_3$H$_{8}$O$_3$) have much higher $\Pi$ ratios that both correspond to $R\simeq 0.5$. This estimation suggests that the phase diagram cannot be described as one-dimensional for the hydrogen-bonded systems, aligning with experimental results.

In summary, the findings of Ref. \onlinecite{Gundermann2011} directly motivate a temporal coarse-graining approach, where the density scaling exponent $\gamma$ can be independently computed from frequency-dependent response functions. Thus, a systematic temporal coarse-graining from first principles is expected to facilitate the understanding of isomorph theory for complex liquids currently accessible only through experiments. 

\subsection{Coarse-graining in space}
In addition to coarse-graining in the temporal domain, one can also consider coarse-graining in space (or \textit{spatial coarse-graining}) by constructing a reduced configurational representation of the complex molecular system in question. For OTP, this specific coarse-graining scheme is inspired by the widely adopted Lewis-Wahnstr\"{o}m model \cite{wahnstrom1993molecular,lewis1993relaxation,lewis1994rotational,Lewis1994,wahnstrom1997translational}. This model consists of three CG sites connected via fixed bonds and interacting through a Lennard-Jones interaction to represent the OTP molecules in an \textit{ad hoc} reduced form. Despite its simplicity and limitations, the Lewis-Wahnstr\"{o}m model can reasonably simulate the chemical and physical behavior of OTP at an efficient computational cost \cite{rinaldi2001dynamics,lombardo2006computational,Pedersen2011}. Notably, as expected from the fixed bond length, the Lewis-Wahnstr\"{o}m model exhibits strong virial-potential energy correlations and obeys isomorph theory predictions \cite{Schroder2009,Ingebrigtsen2012}. However, the interaction parameters for the Lewis-Wahnstr\"{o}m model are not directly parametrized from atomistic OTP energetics, and hence there is no clear microscopic evidence that the hidden scale invariance identified using the \textit{ad hoc} three-site model is representative of atomistic OTP molecules. 

Unlike the \textit{ad hoc} representation, bottom-up spatial coarse-graining approaches can systematically construct a reduced model that faithfully reproduces important microscopic correlations \cite{muller2002coarse,voth2008coarse,peter2009multiscale,noid2013perspective,brini2013systematic,jin2022bottom}. Furthermore, based on fine-grained (fully atomistic) simulations, a bottom-up spatial CG method can answer the following questions. Is the Lewis-Wahnstr\"{o}m model a microscopically consistent representation of OTP molecules? If not, to what extent does this \textit{ad hoc} model accurately capture microscopic correlations at the reduced resolution? Chemical intuition suggests that bonds between linked phenyl rings in OTP should fluctuate, indicating that the underlying assumption of the Lewis-Wahnstr\"{o}m model might be incorrect. In this regard, constructing a three-site spatial CG model using the atomistic OTP trajectory can unravel this ambiguity. 

Due to the expected flexibility of intramolecular bonds in OTP, a single-site (center-of-mass) representation would be the optimal resolution for tracing out a one-dimensional phase diagram because it eliminates directional intramolecular interactions. As such, this spatial renormalization allows us to estimate the virial-potential correlation of the underlying molecule. Eventually, the single-site CG representation can provide insights into density scaling in the following manner: (1) how does the scaling exponent behave at the reduced resolution? By removing unnecessary degrees of freedom, one could apply ideas from the isomorph theory. (2) The renormalized interaction profile can be examined at a single-site resolution to understand the microscopically determined interaction. Based on the obtained CG interaction, we will examine whether it can be approximated as analytical interactions, e.g., IPL, to establish a systematic connection to density scaling. Our overall aim is to link both the density scaling and excess entropy scaling of complex molecular liquids using this systematic spatial coarse-graining approach, as well as temporal coarse-graining.
\begin{figure}
\includegraphics[width=0.9\columnwidth]{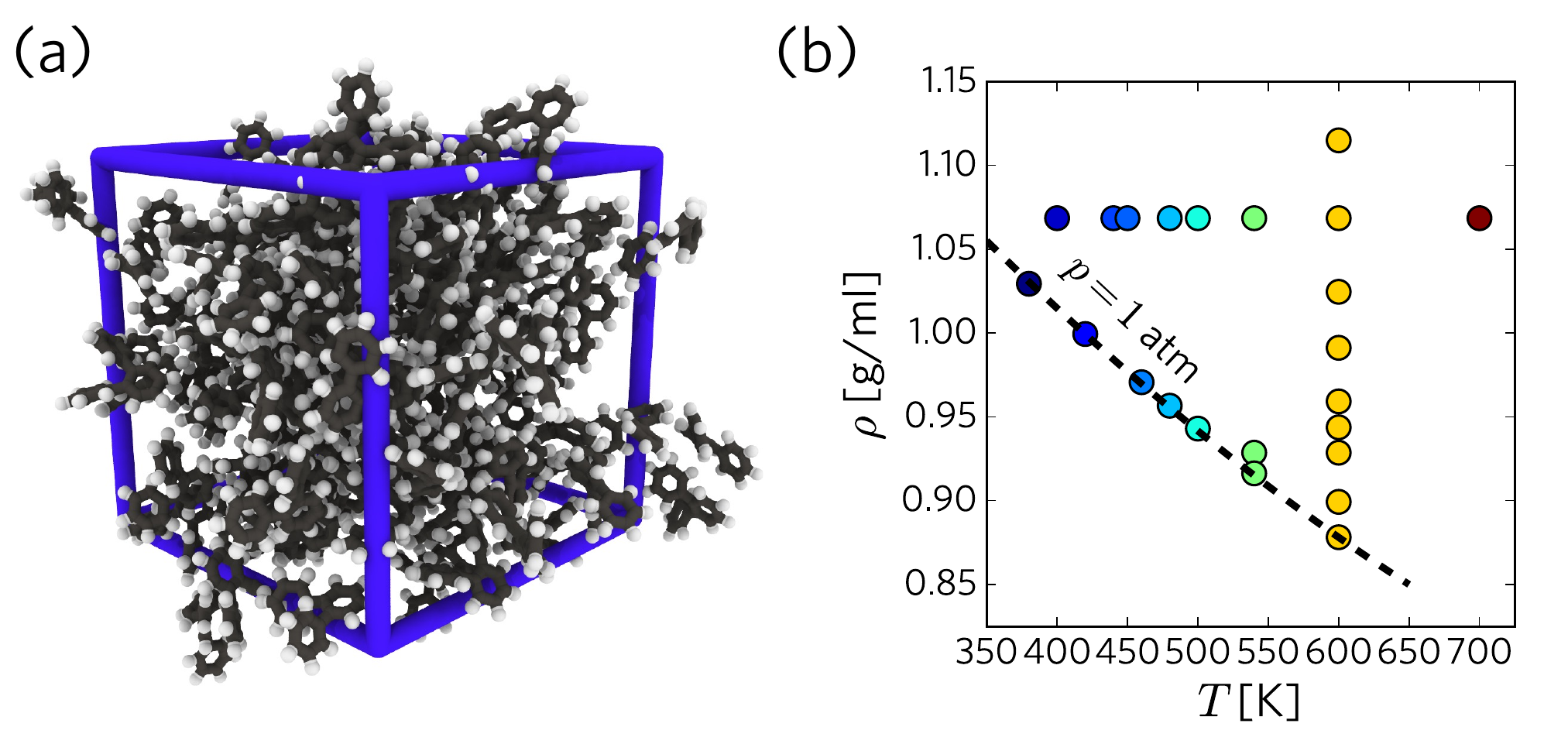}
\caption{\label{fig:OTP-simulation} Atomistic MD simulations of the OTP system. (a) Snapshot from an atomistic simulation of 125 OTP molecules. The blue lines indicate the periodic boundaries of the simulation box. (b) Thermodynamic state points investigated in this work. The dashed line indicates state points of ambient pressure ($1$ atm).} 
\end{figure}
\subsection{Microscopic Reference: Atomistic Simulations}\label{subsec:MD}
In order to perform coarse-graining in space and time, we first performed an all-atom simulation of 125 OTP molecules \cite{Greet1967, Richert2005, Takahara1999, Casalini2016, Mapes2006}, see Supplemental Material (SM) for details \cite{supp}. Figure\ \ref{fig:OTP-simulation}(a) depicts a snapshot of the OTP system modeled  at the fully atomistic resolution, and Fig. \ref{fig:OTP-simulation}(b) illustrates the explored state points to investigate density scaling. Various thermodynamic state points of OTP were investigated, where the temperature ranges from 380 to 700 K and the (mass) density from 0.878 to 1.115 g/cm$^3$ (corresponding to the box lengths from 37.899 to 35 \angstrom), falling between the experimental conditions studied for OTP. At each state point, constant temperature molecular dynamics simulations \cite{GronbechJensen2014, GronbechJensen2014b} were carried out for durations between 70 and 400 ns using the {\tt{LAMMPS}} software package (the ``29 Sep 2021'' version) \cite{Thompson2022}.

\section{Coarse-graining in time}\label{sec:Coarse-graining-in-time}
\subsection{Theory of Temporal Coarse-Graining}

The central goal of \emph{coarse-graining in time} is to average out fast degrees of freedom while retaining the slow degrees of freedom. We assume the system is at a sufficiently low temperature, where intramolecular degrees of freedom, such as bond vibrations, are considerably faster than molecular motions, e.g., translational motions. We consider temporal averaging characterized by a time scale, $\tau$, or the inverse angular frequency, $\omega^{-1}$, chosen to be faster than the intermolecular dynamics but slower than intramolecular motions. In general, the optimal $\tau$ and $\omega$ depend on the specific observable. In this section, we will introduce several temporal coarse-graining methodologies and show that these capture nearly the same strong correlation nature for OTP. 

\subsection{Time-Averaging Approach}
Arguably, the most straightforward method for implementing temporal coarse-graining would be to perform a time average of the potential energy $U(t)$ and the virial $W(t)$. This \emph{time-averaging} approach is motivated by conventional experimental settings, where measuring fluctuations with a probe can be thought of as an effective ``low-pass'' filter. This observable corresponds to the temporal CG quantity $\bar f(t; \tau)$, which can be expressed as a convolution
\begin{eqnarray}\label{eq:convolution}
\bar f(t; \tau) 
&=& \int_{-\infty}^\infty f(t')w(t-t';\tau)dt', 
\end{eqnarray}
where $w(t; \tau)$ is a kernel with the property $\int_{-\infty}^{\infty}w(t; \tau) dt=1$ that describes how the averaging is distributed in time (not to be confused with the angular frequency $\omega$). $\bar{f}$ is associated with slow fluctuations if the $w$ function has a significant width, usually quantified by a large value of the $\tau$ parameter. The quantity $\bar f(t; \tau)$ is relevant to experimental measurements, where $f(t)$ might not be readily available. We implemented a discrete analog of Eq. (\ref{eq:convolution}) using the SciPy Python package \cite{SciPy} that is publicly available in Ref. \onlinecite{zenodo}.

\begin{figure}
\includegraphics[width=1.0\columnwidth]{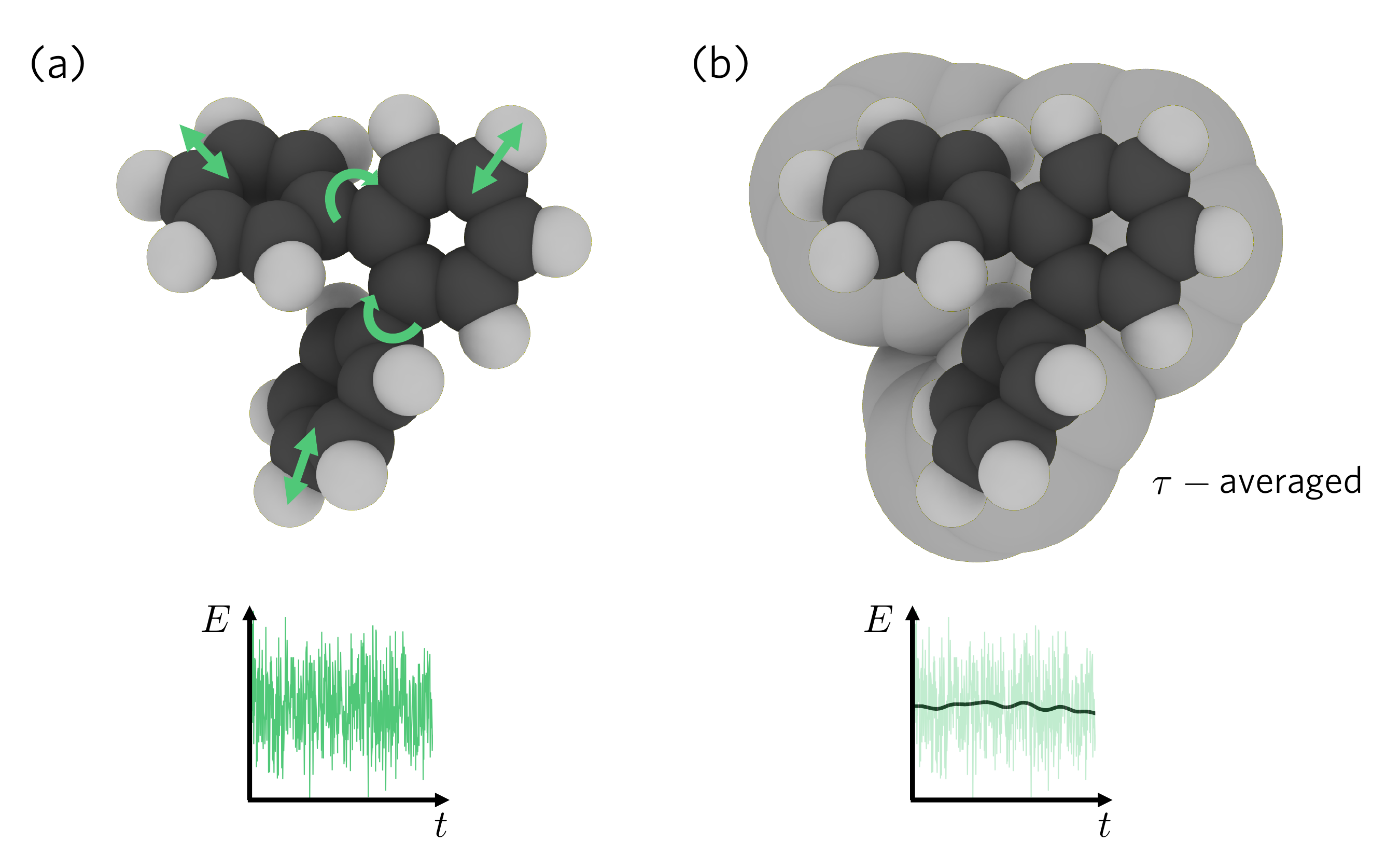}
\caption{\label{fig:temporalCG} Schematic description of the proposed temporal coarse-graining: (a) Fully atomistic simulation trajectories give energetics with frequent fluctuations (green lines) due to intramolecular interactions and motions (green arrows). (b) By temporal coarse-graining the energetics over the characteristic time $\tau$, one can focus on the slow fluctuations (black lines) due to the intermolecular interactions.}
\end{figure}

The $w$ is typically an exponential decay, a Gaussian function, or the Hann function. In this work, we adopt the ``boxcar'' average, where $\bar f(t, \tau)$ is computed by averaging values from $f(t)$ to $f(t+\tau)$ with equal weight. The temporal coarse-graining here is achieved using a rectangular window as the $w$ function, defined as
\begin{equation}
w(t; \tau)=\begin{cases}
    0  & \text{for } t<0,\\
    1/\tau & \text{for } 0\leq t\leq t+\tau,\\
    0 \quad &\text{for } t>t+\tau.\\
     \end{cases}
\end{equation}
As $w(t;\tau)\rightarrow\delta(t)$ for $\tau\rightarrow0$, the direct signal $f(t)$ is recovered in the limit $\tau\to0$.

\begin{figure*}
\includegraphics[width=1.6\columnwidth]{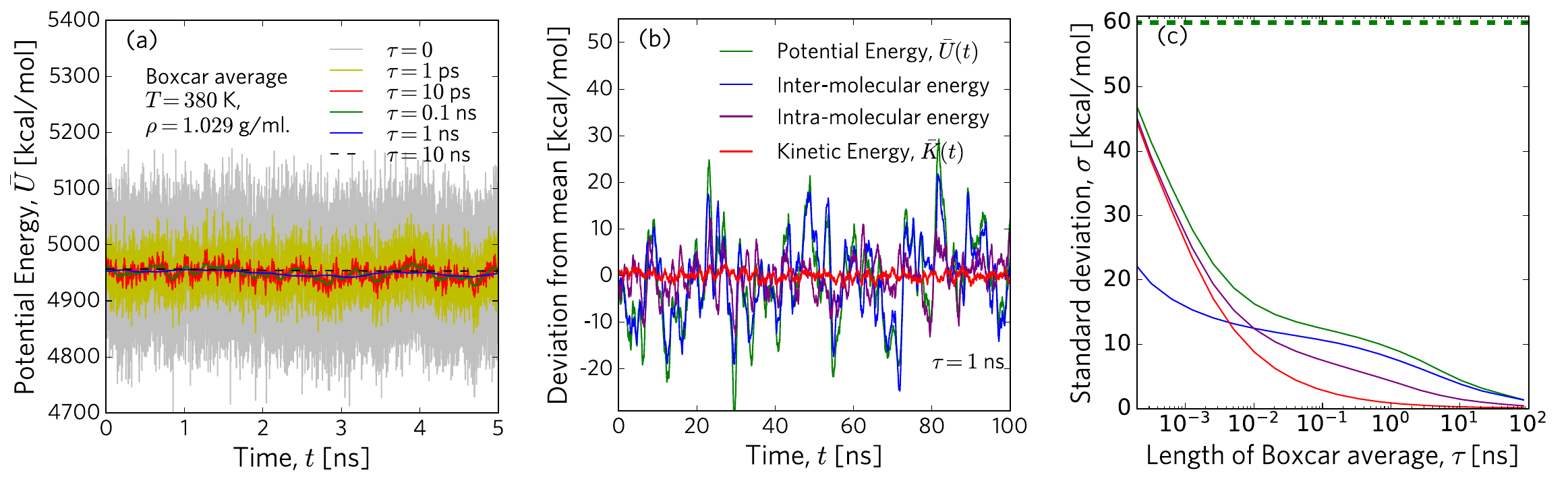}
\caption{\label{fig:boxcar} Temporal coarse-graining by the time-averaging approach. (a) Boxcar average of the potential energy, $\bar U(t; \tau)$, in a simulation of 125 OTP molecules at $T=380$ K and $\rho=1.029$ g/ml (1 atm) for different choices of the characteristic time scale, $\tau$ (see legend). For clarity, only 5 ns of the 468 ns simulation is shown. (b) Various time-averaged observables using $\tau=1$ ns, including the time-averaged potential energy (green); intermolecular (blue) and intramolecular (purple) energies decomposed using Eq. (S1) \cite{supp}; and kinetic energy (red). (c) Standard deviations of the time-averaged energetics from panel (b) of the potential, intermolecular, intramolecular, and kinetic energies, respectively.}
\end{figure*}

Figure \ref{fig:boxcar}(a) demonstrates the boxcar average of potential energy fluctuations with averaging lengths ranging from $\tau=1$ ps to $\tau=10$ ns at $T = 380$ K and $\rho = 1.029$ g/ml, corresponding to ambient pressure. As expected, the fluctuations decrease as $\tau$ increases. Figure \ref{fig:boxcar}(b) displays the time-averaged potential energy, the average of the intramolecular contribution, and of the intermolecular contribution with $\tau=1$ ns, representing slow fluctuations. Notably, using $\tau=1$ ns, the kinetic energy fluctuations have diminished, whereas the fluctuations of intermolecular energy remain significant. These data show that upon temporal coarse-graining, the slow fluctuations are representative of the intermolecular contributions. 

In order to systematically investigate the impact of $\tau$, Fig.\ \ref{fig:boxcar}(c) presents the standard deviation as a function of $\tau$, revealing that slow energetic fluctuations are observed when the width of the average is on the order of a nanosecond. For some state points, separating fast and slow fluctuations is not possible, especially when the structural relaxation time is significantly faster than one nanosecond. Moreover, in scenarios of low temperatures or high densities, obtaining good statistics for slow fluctuations can be challenging as the dynamics become sluggish. In this study, we find that a relaxation time of one nanosecond is an appropriate characteristic time for temporal coarse-graining for OTP. Table \ref{tab:boxcar_gammas} lists additional $\tau$ values determined across various state points through the coarse-graining process [Fig. \ref{fig:CUU_time_correlation}(a)]. 

\begin{figure}
\includegraphics[width=0.45\columnwidth]{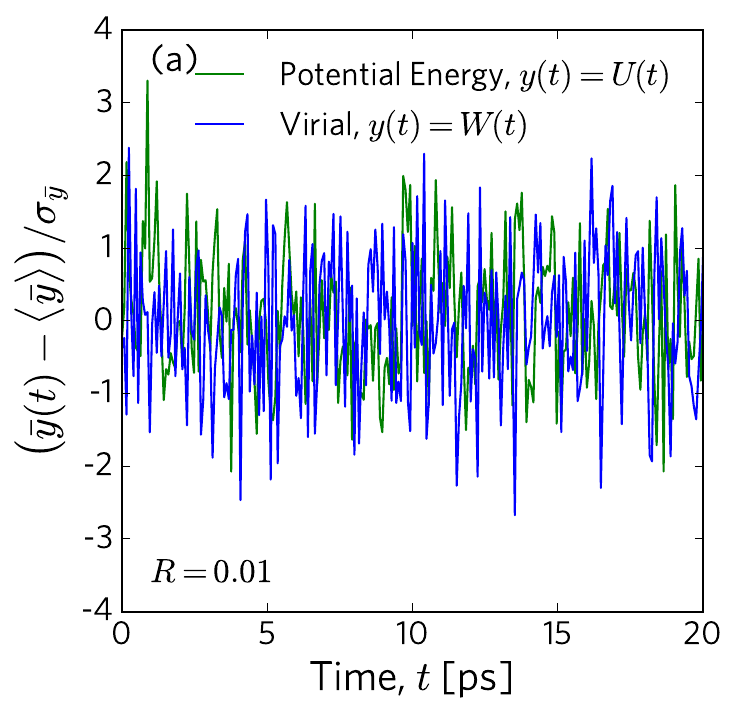}
\includegraphics[width=0.45\columnwidth]{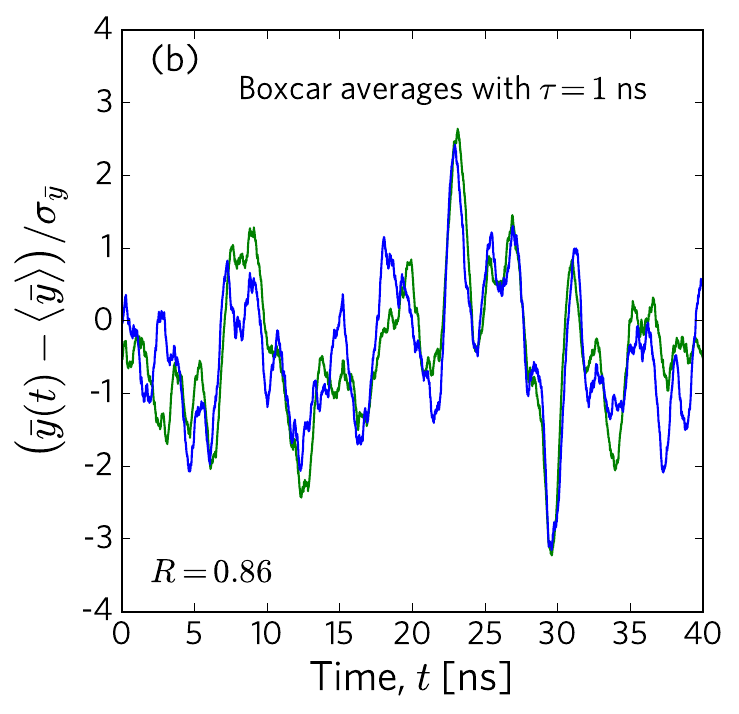}
\caption{\label{fig:boxcar_virial_energy} $WU$ fluctuations from time-averaged CG OTP. (a) Normalized times series, $(y(t)-\langle y \rangle)/\sigma_y$ (where $\sigma_y$ is the standard deviation), of potential energy (green) and viral (blue). The correlation is noticeably weak with a Pearson correlation coefficient $R=0.01$. (b) Boxcar averaged energy (green) and virial (blue) fluctuations. The correlation is strong, quantified by $R=0.86$. Note that the time-axis on this panel is in units of nanoseconds, while it is in picoseconds in the first panel.}
\end{figure}

\begin{figure}
\includegraphics[width=0.45\columnwidth]{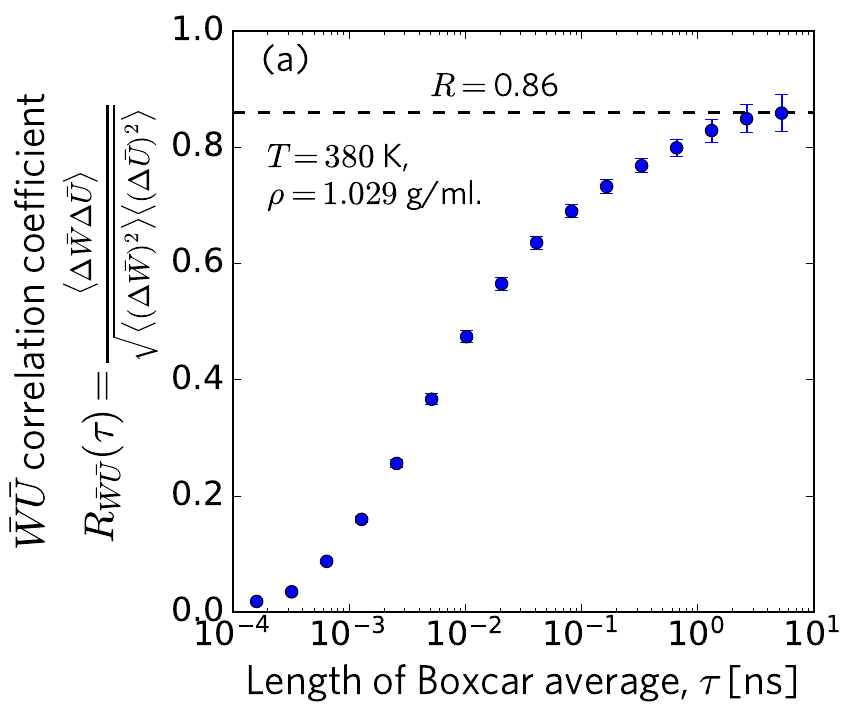}
\includegraphics[width=0.43\columnwidth]{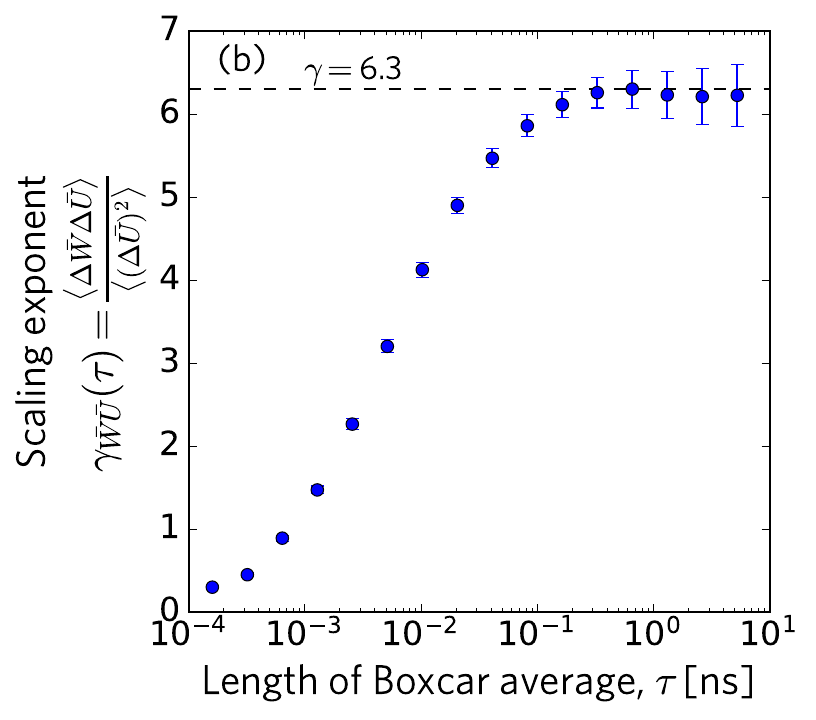}
\caption{\label{fig:boxcar-density-scaling} Correlation coefficients from time-averaged CG OTP. (a) The Pearson correlation coefficient of the boxcar averaged potential energy ($\bar U$) and virial ($\bar W$), as a function of the length of the boxcar average, $R_{\bar W\bar U}(\tau)=\frac{\langle \Delta \bar W\Delta \bar U\rangle}{\sqrt{\langle (\Delta \bar W)^2\rangle\langle (\Delta \bar U)^2\rangle}}$. Strong correlations of the slow $\bar U$ and $\bar W$ fluctuations are evident as $R\to0.86$ for $\tau\to\tau_{\alpha}$ (black dashed line). Error bars (67\% confidence intervals where the standard score $Z=1$) throughout this work are estimated by dividing the trajectories into eight statistically independent blocks \cite{Flyvbjerg1989}.
(b) Density scaling exponent ($\gamma$) estimated from $\bar U$ and virial $\bar W$. We find that $\gamma\to6.3$ for $\tau\to\tau_{\alpha}$ (black dashed line).}
\end{figure}

Next, we investigate whether the CG energy landscape exhibits strong correlations and eventually scale invariance, as depicted in Fig.\ \ref{fig:energy_landscape}. Drawing inspiration from isomorph theory for point particles, we expect strong correlations in the slow fluctuations between the potential energy and the virial: While the instantaneous fluctuations of potential energy and virial themselves are virtually uncorrelated [Fig.\ \ref{fig:boxcar_virial_energy}(a)], the slow fluctuation are correlated [Fig.\ \ref{fig:boxcar_virial_energy}(b)]. Therefore, the Pearson correlation coefficient,
\begin{equation}
    R_{\bar U \bar W}(\tau) = \frac{\langle \Delta \bar W \Delta \bar U \rangle}{\sqrt{\langle (\Delta \bar W)^2\rangle\langle (\Delta \bar U)^2\rangle}}
\end{equation}
should be close to unity, in which $\Delta \bar W$ and $\Delta \bar U$ are the temporal CG observables defined by Eq. (\ref{eq:convolution}). As depicted in Fig. \ref{fig:boxcar-density-scaling}(a), $R_{\bar U \bar W}(\tau)$ approaches 0.86 for $\tau>$ 1 ns, validating strong correlations. This suggests that the slowly fluctuating part of the potential energy landscape is scale-invariant, aligning with the illustration in Fig.\ \ref{fig:energy_landscape}. Building upon this observation, we proceed to compute the density scaling exponent from temporal coarse-graining using 
\begin{equation}
    \gamma_{\bar U\bar W}(\tau) = \frac{\langle \Delta \bar W \Delta \bar U \rangle}{\langle (\Delta \bar U)^2 \rangle}.
\end{equation}
As demonstrated in Fig. \ref{fig:boxcar-density-scaling}, the exponent converges to 6.3(3). Extending this approach to other state points with similar structural relaxation times, the averaged estimates for scaling exponents reported in Table \ref{tab:boxcar_gammas} consistently yield $\gamma$ values of 6.3 within the statistical uncertainty.  

In summary, we have demonstrated that by directly performing time averaging of fluctuating atomistic virial and potential energies, an average value of $\gamma=6.3$ for atomistic OTP is obtained, indicating a strongly correlating nature of the slow degrees of freedom that previously was challenging to assess.

Before examining whether this $\gamma$ value correctly encodes the experimentally observed one-dimensional phase diagram of OTP, in the remainder of this section, we investigate whether the time-averaging approach represents the correct temporal CG characteristic of OTP using alternative approaches inspired by experimental observations.

\begin{table}
    \centering
    \caption{State points where the density scaling exponent could be accurately determined from the slow $\bar W(t)$-$\bar U(t)$ fluctuations using a boxcar average, compare Fig.\ \ref{fig:boxcar-density-scaling}. The structural relaxation time $\tau_\alpha$ is estimated as shown in Fig.\ \ref{fig:CUU_time_correlation}(a).}
    \label{tab:boxcar_gammas}
    \begin{tabular}{
        S[table-format=3.0] 
        S[table-format=1.3] 
        S[table-format=1.1] 
        S[table-format=1.1] 
    }
    \toprule
    \hline
    {$T$ [K]} & {$\rho$ [g/ml]} & {$\tau_\alpha$ [ns]} & {$\gamma$} \\
    \midrule
    \hline
    360 & 1.045 & 1.0 & 6.1 \\
    380 & 1.029 & 0.5 & 6.3 \\
    400 & 1.060 & 2.0 & 6.0 \\
    400 & 1.014 & 0.1 & 6.2 \\
    500 & 1.096 & 1.0 & 7.0 \\
    500 & 1.078 & 0.4 & 6.3 \\
    \bottomrule
    \hline
    \end{tabular}
\end{table}

\subsection{Time Correlation Function Approach}

As we are interested in the time series of fluctuations, an alternative approach to performing temporal coarse-graining can be derived from the time correlation formalism. The time correlation function between two signals $f$ and $g$ with a lag time $\tau$ is defined as
\begin{equation} \label{eq:timecorrel}
    C_{fg}(\tau) = \overline{f(\tau)g(0)} = \lim_{T_0 \rightarrow \infty} \frac{1}{T_0} \int_{0}^{T_0} f(t+\tau)g(t)dt.
\end{equation}
Efficient computation of Eq. (\ref{eq:timecorrel}) is possible through the cross-correlation theorem combined with the Fast Fourier Transform (FFT) algorithm, as described in Ref. \onlinecite{numerical_recipes}. To briefly outline the standard procedure we followed, let $\mathcal{F}[y(t)]=\int_{-\infty}^{\infty} y(t)\exp(-i\omega t)dt$ be the Fourier transform and $\mathcal{F}^{-1}[Y(\omega)]=\int_{-\infty}^{\infty} Y(\omega)\exp(i\omega t)d\omega$ be the inverse Fourier transform. The time correlation function $C_{fg}$ can then be expressed through the inverse Fourier transform of the convolution in the frequency domain:
\begin{equation} \label{eq:Cfg}
    C_{fg}(\tau) = \mathcal{F}^{-1}[F^*(\omega)G(\omega)],
\end{equation}
where $F(\omega)=\mathcal{F}[f(t)]$, $G(\omega)=\mathcal{F}[g(t)]$, and $(\ldots)^*$ represents the complex conjugate. From its definition, Eq. (\ref{eq:timecorrel}) represents slow fluctuations as $\tau$ increases. 

\begin{figure*}
\includegraphics[width=1.6\columnwidth]{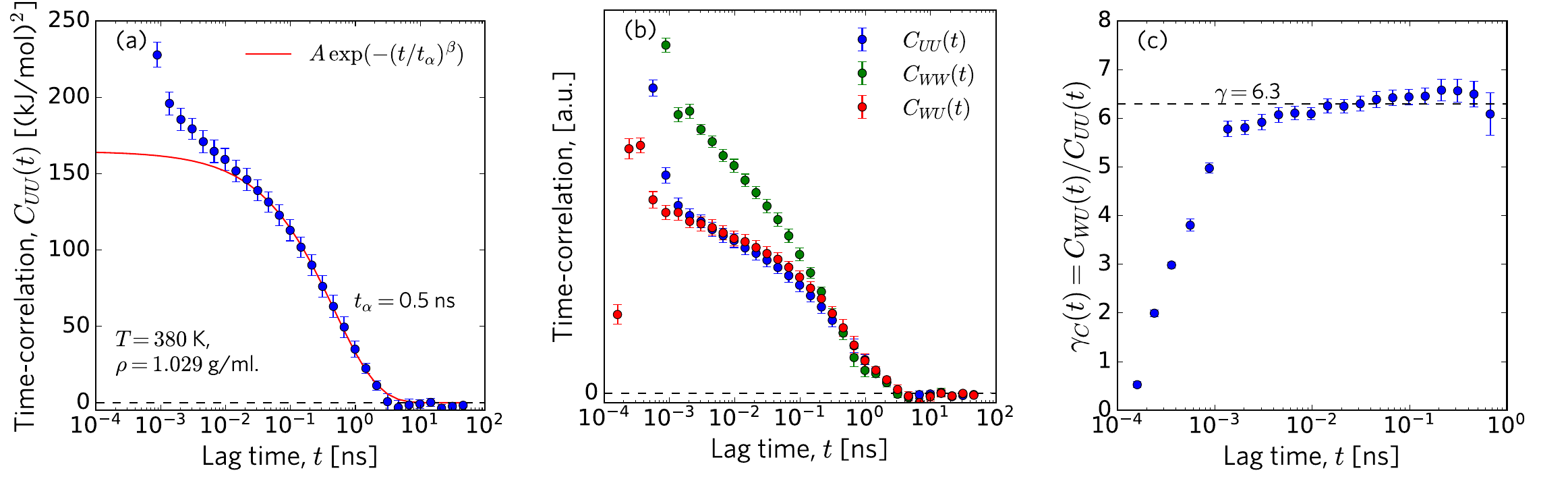}
\caption{\label{fig:CUU_time_correlation} Temporal coarse-graining by the time correlation function approach. (a) 
Autocorrelation functions of the potential energy fluctuations, $C_{UU}(t)$. The long-time decorrelation of energy fluctuations is used to estimate the structural relaxation time of the system to $\tau_\alpha=0.5$ ns by fitting to a stretched exponential $A\cdot\exp\left[{-(t/\tau_\alpha)^\beta}\right]$ (red line), where $A=170$ (kJ/mol)$^2$ and  $\beta=0.62$. (b) Comparison of the time correlation functions $C_{UU}(t)$, $C_{WW}(t)$, and $C_{UW}(t)$. The vertical axis has been scaled so that long-time relaxation overlaps with the expected correlations from slow fluctuations. (c) The density scaling exponent estimated from the time correlation functions, see Eq.\ (\ref{eq:gamma_time_correlation}), is consistent with the results of Fig.\ \ref{fig:boxcar-density-scaling}.}
\end{figure*}

In order to assess the CG virial-potential energy correlation using the time correlation formalism, Fig. \ref{fig:CUU_time_correlation}(a) illustrates the time correlation function of the potential energy with itself over time, $C_{UU}(t)$---the autocorrelation function, at the same thermodynamic state point as investigated in the previous subsection using time-averaging (Figs. \ref{fig:boxcar}--\ref{fig:boxcar-density-scaling}): $380$ K; $1.029$ g/ml; 1 atm. The autocorrelation function depicted in Fig. \ref{fig:CUU_time_correlation}(a) demonstrates the multi-step relaxation nature of the potential energy surfaces. As the terminal relaxation is associated with important molecular translations and rotations, we estimated the structural relaxation time underlying this terminal relaxation by fitting to a stretched exponential, $A \cdot \exp\left[-(t/\tau_\alpha)^\beta\right]$. At this specific state point, the structural relaxation time was determined to be $\tau_\alpha=0.5$ ns, which is within the same order of magnitude as the time scale obtained from the time-averaging method [cf. Fig.\ \ref{fig:boxcar}(c) and Fig.\ \ref{fig:CUU_time_correlation}(a)]. 

As we apply the time correlation formalism to other fluctuating quantities, a strong correlation of the slow energy-virial fluctuations becomes evident. In Fig. \ref{fig:CUU_time_correlation}(b), we observe that the long-time (slow) relaxation behavior of $C_{UU}(t)$, $C_{WW}(t),$ and $C_{UW}(t)$ coincide near the terminal relaxation time. This feature suggests strong correlations between the virial and the potential energy at this characteristic time. Based on these findings, we computed the scaling exponent estimated from the time correlation function [Fig. \ref{fig:CUU_time_correlation}(c)]
\begin{equation}\label{eq:gamma_time_correlation}
    \gamma_C(t) = \frac{C_{UW}(t)}{C_{UU}(t)},
\end{equation}
which provides a value consistent with that of the time-averaging approach [Fig. \ref{fig:boxcar-density-scaling}(b)],  $\gamma = 6.3$. 

\subsection{Frequency-Dependent Response Approach}
Our final temporal CG approach is inspired by studies of the frequency-dependent Prigogine-Defay ratio for understanding density scaling \cite{Ellegaard2007, Pedersen2008b, phd_thesis, Gundermann2011}. This concept can be formally derived from the fluctuation-dissipation theorem, which links the power spectrum of equilibrium fluctuations to frequency-dependent linear response functions. Generally, a frequency-dependent response function related to the variables $f$ and $g$ can be computed as a Fourier-Laplace transform
\begin{equation} \label{eq:flucdiss}
    \mu_{fg}(\omega) = \int_0^{\infty} \dot C_{fg}(t)\exp(-i\omega t)dt,
\end{equation}
in which $\dot C_{fg}(t) = dC_{fg}(t)/dt$. Since $\dot C_{fg}$ approaches zero when $t\rightarrow0$, the integral in Eq. (\ref{eq:flucdiss}) can be treated as a conventional Fourier transform when $\dot C_{fg}(t)$ is prescribed to be zero at negative times. Consequently, this quantity can be efficiently computed using the FFT algorithm. For example, the frequency-dependent heat capacity can be estimated from $c_{V}(\omega)=-\mu_{UU}(\omega)/k_BT^2$ \cite{nielsen1996fluctuation}.

Having established the relationship between the frequency-dependent response function and the power spectrum of fluctuations, we now evaluate the correlation between the virial and the potential energy fluctuations. At low angular frequencies corresponding to long time scales, this can be done by assessing $\mu_{UU}(\omega)$, $\mu_{WU}(\omega)$, and $\mu_{WW}(\omega)$, as depicted in Fig.\ \ref{fig:response}(a). From these frequency-dependent response functions, the generalized correlation coefficient between two signals (which is also frequency-dependent) can be defined in a manner analogous to the frequency-dependent Prigogine-Defay ratio \cite{Ellegaard2007, Pedersen2008b, phd_thesis, Gundermann2011}:
\begin{eqnarray} \label{eq:prigogine}
    R_\mu(\omega) &\equiv& \frac{\mathfrak{I}\mu_{UW}(\omega)}{\sqrt{\mathfrak{I}\mu_{WW}(\omega)\mathfrak{I}\mu_{UU}(\omega)}}.
\end{eqnarray}
In Eq. (\ref{eq:prigogine}), $\mathfrak{I}(\ldots)$ denotes the imaginary part of the complex response functions. In practice, $R_\mu(\omega)$ is rather noisy, yet it confirms the existence of strong virial-potential energy fluctuations. From Eq. (\ref{eq:prigogine}), the corresponding frequency-dependent scaling exponent can be defined as
\begin{equation}
    \gamma(\omega) = \frac{\mathfrak{I}\mu_{WU}(\omega)}{\mathfrak{I}\mu_{UU}(\omega)},
\end{equation}
which can still be estimated from the noisy response functions. Figure \ref{fig:response}(b) further confirms the validity of our approach, consistently yielding $\gamma=6.3$ as with the previous two approaches.

\subsection{Summary}
We have introduced three methods for coarse-graining the fast intramolecular degrees of freedom in OTP molecules over a temporal domain. By effectively retaining the slow degrees of freedom through temporal coarse-graining, we have shown that this temporal CG approach allows for uncovering strong correlations between the virial and potential energy of OTP that are challenging to assess at a fully atomistic resolution. Specifically, our temporal coarse-graining approaches include (1) direct time-averaging of the fluctuating virial and potential energy, (2) indirect coarse-graining of the time correlation formalism in conjunction with Fourier transformation, and (3) utilization of a frequency-dependent response to coarse-grain the power spectrum of fluctuations via the fluctuation-dissipation theorem. 

\begin{figure}
\includegraphics[width=0.45\columnwidth]{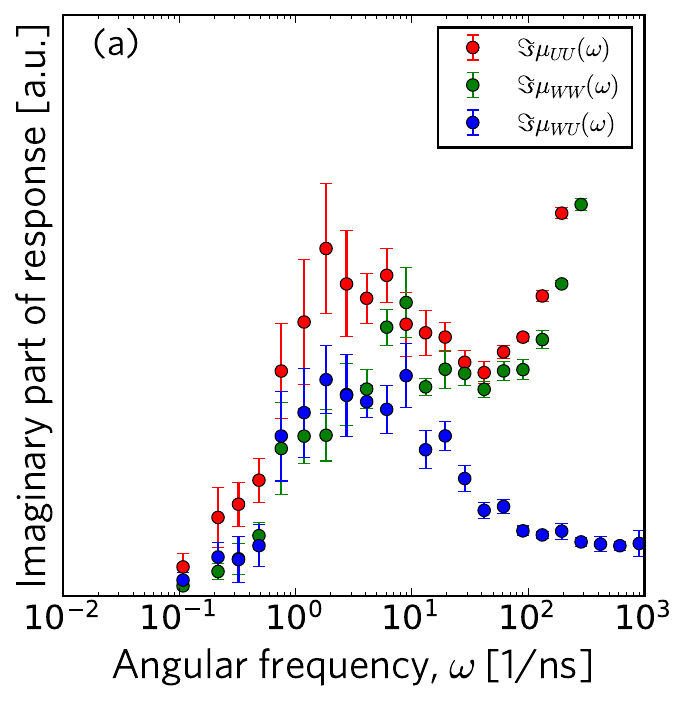}
\includegraphics[width=0.48\columnwidth]{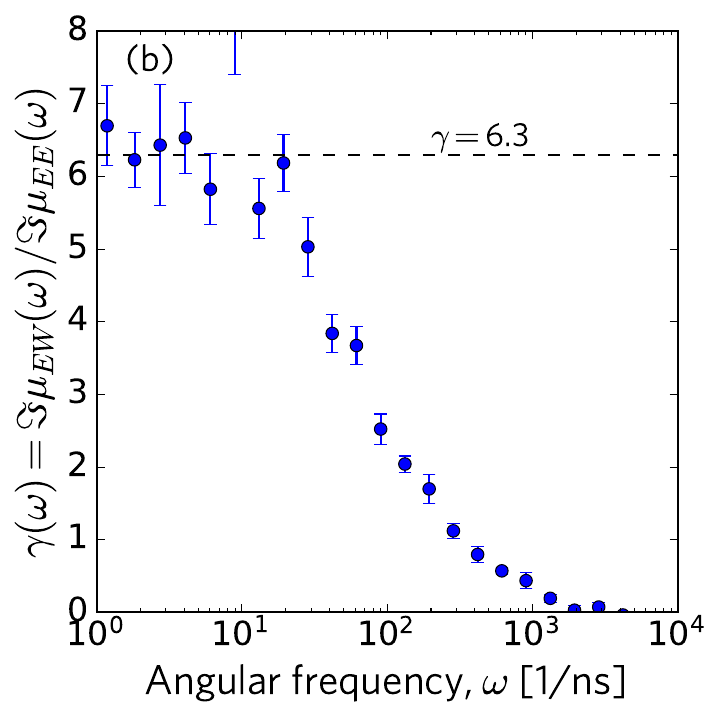}
\caption{\label{fig:response} Temporal coarse-graining by the frequency-dependent linear response function approach. (a) Imaginary parts of frequency-dependent response functions. (b) The frequency-dependent density scaling exponent.}
\end{figure}
We have implemented these three different temporal CG approaches for OTP across various temperature and density conditions. These distinct approaches yield a scaling exponent near $\gamma=6.3$ at various thermodynamic state points, with values ranging from 6.23 to 6.49 as shown in Figs. \ref{fig:boxcar-density-scaling}(b), \ref{fig:CUU_time_correlation}(c), and \ref{fig:response}(b). This agreement indicates that, while the approaches are derived differently, the underlying physical picture of the strong correlation in OTP is invariant. Nevertheless, among the different approaches, we note that the time-averaging method could be easily extended to other complex molecules as it provides better statistics and is easier to implement than the other two approaches. As a prototypical demonstration of the applicability of the time-averaging approach, we extend the temporal CG approach to a single-component Lennard-Jones system with both repulsion and attraction by recapitulating the strong correlations, see SM (Sec. V). Altogether, temporal coarse-graining reveals a hidden scale invariance of the slow fluctuations in the OTP energy landscape [Fig. \ref{fig:boxcar_virial_energy}(b)], showing that $U({\bf r}^n)$ in OTP resembles the landscape hypothesized in Fig.\ \ref{fig:energy_landscape}. 

\section{Coarse-Graining in Space}\label{sec:Coarse-graining-in-space}
\subsection{Theory of Spatial Coarse-Graining}
\begin{figure}
\includegraphics[width=0.9\columnwidth]{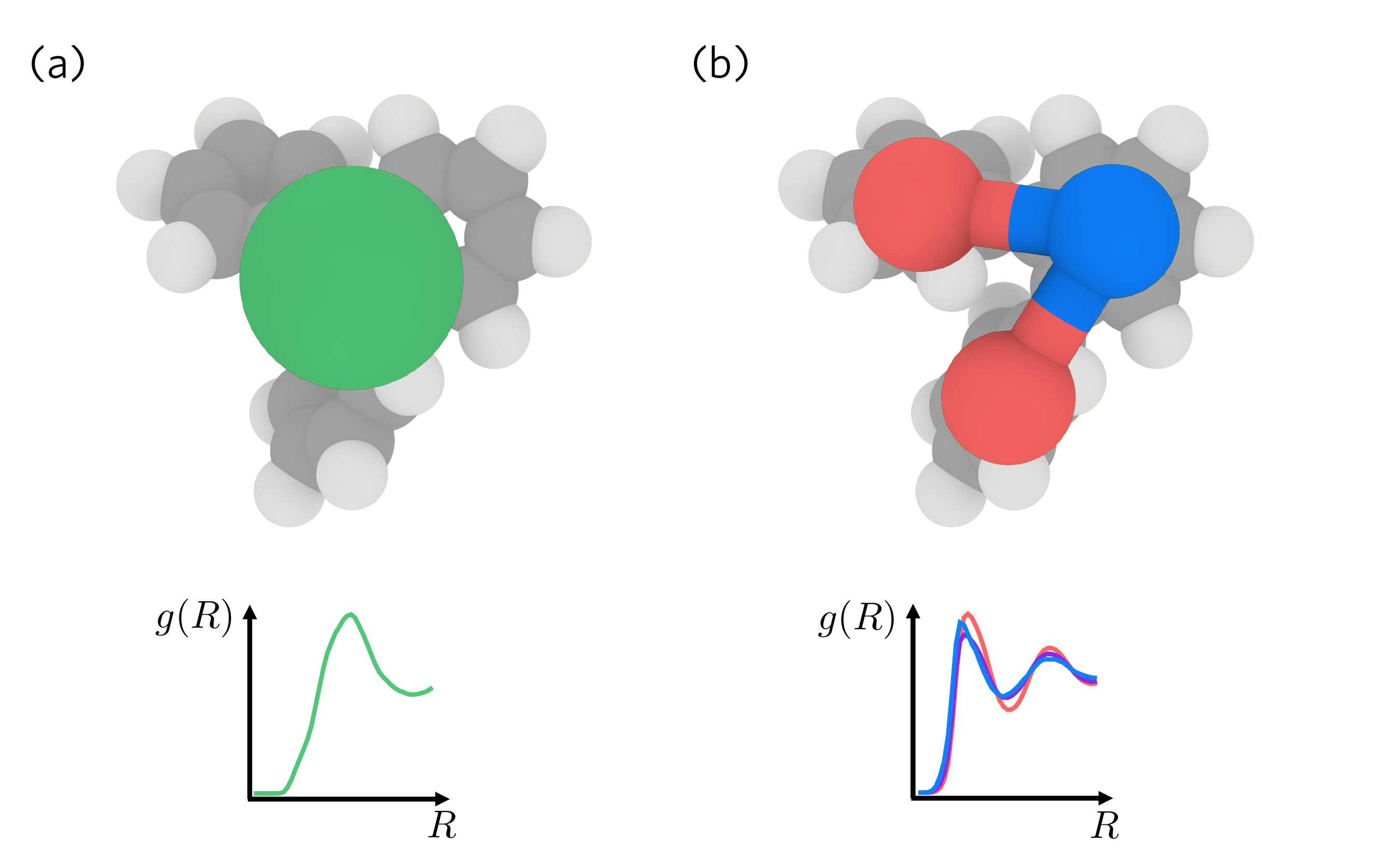}
\caption{\label{fig:spatialCG} Schematic of the proposed spatial coarse-graining with two different mapping schemes and the corresponding radial distribution functions : (a) Single-site CG model at the center-of-mass level. The single-site resolution integrates the intramolecular degrees of freedom and is thus suitable for investigating hidden scale invariance. (b) Three-site CG model. This resolution is adopted by the \textit{ad hoc} Lewis-Wahnstr\"{o}m model but lacks a correspondence to microscopic physics.}
\end{figure}

While coarse-graining in time entails convoluting the observables over time, \textit{coarse-graining in space} aims to simplify the molecular system itself \cite{muller2002coarse,voth2008coarse,peter2009multiscale,noid2013perspective,brini2013systematic,jin2022bottom}. For OTP, the primary objective is to design a single-site CG model composed of $N$ particles ($N\ll n$) free of intramolecular interactions. The desired spatial CG model should accurately capture the effective correlations of atomistic OTP molecules at the center-of-mass level [Fig. \ref{fig:spatialCG}(a)]. Bottom-up coarse-graining is the optimal strategy for performing coarse-graining in space, as it aims to faithfully recapitulate the microscopic (atomistic) correlations at the spatial CG level \cite{noid2013perspective,jin2022bottom}.

Consider a spatial coarse-graining mapping operator, denoted as $M_\mathbf{R}^N$, which transforms the fine-grained (atomistic) configurations $\mathbf{r}^n$ into the CG configurations $\mathbf{R}^N$, i.e., $M_\mathbf{R}^N:\mathbf{r}^n \rightarrow \mathbf{R}^N$. The CG model correctly approximates the fine-grained (FG) reference when the following consistency condition regarding the equilibrium probability distributions of the FG and CG variables in phase space is satisfied \cite{noid2008multiscale,noid2008multiscale2}: 
\begin{equation}\label{eq:consistency}
    p_\mathrm{CG}({\bf R}^N) = \int \mathrm{d}\mathbf{r}^n \delta\left( M_\mathbf{R}^N\left(\mathbf{r}^n\right)-\mathbf{R}^N\right)p_\mathrm{FG}(\mathbf{r}^n),
\end{equation}
where the delta function enforces that the mapped FG configurations are matched to the CG configurations, which is expressed as a product of delta functions for each CG particle $I$, i.e., $\delta\left( M_\mathbf{R}^N(\mathbf{r}^n)-\mathbf{R}^N\right)=\prod_I^N \left(\delta \left(M_I(\mathbf{r}^n) -\mathbf{R}_I \right)\right)$. 
From the thermodynamic consistency for configurational variables [Eq. (\ref{eq:consistency})], the effective CG interaction $U_\mathrm{CG}$ can be derived as
\begin{align}\label{eq:CGPMF}
    U_\mathrm{CG}=&-k_BT\ln \int \mathrm{d}\mathbf{r}^n \delta\left( M_\mathbf{R}^N(\mathbf{r}^n)-\mathbf{R}^N\right) \nonumber \\ &\times \exp \left( -\frac{u_\mathrm{FG}(\mathbf{r}^n)}{k_BT} \right) + \mathrm{(constant)},
\end{align}
in which $u_\mathrm{FG}(\mathbf{r}^n)$ is the FG interaction potential. It is worth noting that $U_\mathrm{CG}$ in Eq. (\ref{eq:CGPMF}) takes the form of the many-body potential of mean force (PMF) in terms of CG variables \cite{noid2008multiscale,noid2008multiscale2}. Therefore, $U_\mathrm{CG}(\mathbf{R}^N)$ can be interpreted as a free energy quantity and, in general, will vary with the thermodynamic state point. This state point-dependent nature of the CG interactions is referred to as the \textit{transferability issue} in CG modeling \cite{dunn2016van,jin2019understanding,jin2020temperature}.

While Eq. (\ref{eq:CGPMF}) is formally exact such that it yields a thermodynamically consistent CG model, the practical determination of $U_\mathrm{CG}$ through the many-dimensional integration is highly prohibitive in practice. Various optimization schemes have been developed to overcome this challenge. In this work, we employ a multiscale coarse-graining (MS-CG) methodology \cite{noid2008multiscale,noid2008multiscale2} that utilizes force-matching to determine $U_\mathrm{CG}$ \cite{lu2010efficient}. This is achieved by variationally minimizing the force residual $\chi^2[\phi]$, which is the force difference between the microscopic reference force at the CG level, $\mathbf{f}_I(\mathbf{r}^n)$, and the unknown CG force, $\mathbf{F}_I (M_\mathbf{R}^N (\mathbf{r}^n);\phi)$, defined using the CG force field parameter $\phi$:
\begin{equation}\label{eq:forceresidual}
    \chi^2[\phi] = \frac{1}{3N}\left\langle \sum_{I=1}^N |\mathbf{F}_I (M_\mathbf{R}^N (\mathbf{r}^n);\phi)-\mathbf{f}_I(\mathbf{r}^n)|^2 \right\rangle. 
\end{equation}
Thus, $\phi$ can be determined by minimizing Eq. (\ref{eq:forceresidual}). Noid \textit{et al.} have demonstrated that the least-squares solution to Eq. (\ref{eq:forceresidual}) satisfies the consistency relationship given in Eq. (\ref{eq:consistency}) \cite{noid2008multiscale}. This thermodynamically consistent CG model also captures important structural correlations from the atomistic reference: Ref. \onlinecite{noid2007multiscale} established that the force-matching equation [Eq. (\ref{eq:forceresidual})] employing two-body basis sets is a discretized representation of the Yvon-Born-Green theory in liquid physics \cite{theory_of_simple_liquids}. This finding suggests that the MS-CG model can effectively capture up to three-body correlations using two-body basis sets, distinguishing it from other bottom-up methodologies. Therefore, the MS-CG method stands as a robust choice for constructing a CG model of OTP with high structural accuracy.

\subsection{Spatial CG Model of OTP} \label{subsec:CGSpace}
\subsubsection{Effective CG Interaction}
The spatial CG model for OTP was constructed by determining the spatial CG interaction using Eq. (\ref{eq:forceresidual}) at the center-of-mass level. The effective CG interaction follows a pairwise form, and to capture intricate interaction profiles, we introduced the B-spline function to define the pairwise basis sets \cite{lu2010efficient}. Further numerical details are provided in Appendix A. While utilizing B-splines allows for capturing complex interaction profiles that cannot be fully resolved using analytical interaction, it introduces a numerical challenge when estimating the derivatives of interactions. This drawback will be discussed in more detail in Sec. \ref{subsec:representability}, when directly estimating the density scaling exponent $\gamma$ from the CG models. 

The effective pairwise CG interactions of OTP were parametrized by solving Eq. (\ref{eq:FM}) across the thermodynamic state points investigated in Sec. \ref{subsec:MD} spanning temperatures from 380 to 700 K and densities from 0.878 to 1.115 g/ml. 
\begin{figure}
    \includegraphics[width=0.96\columnwidth]{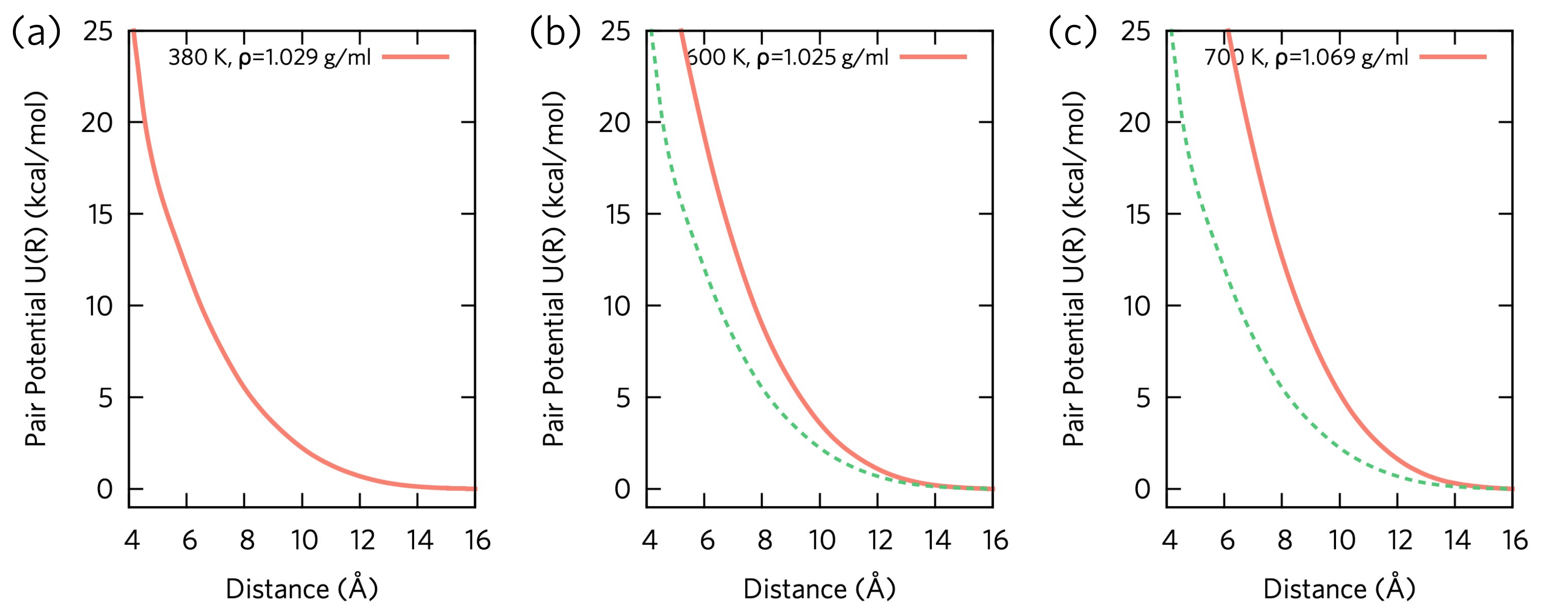}
    \caption{\label{fig:OTP-interaction} Parametrized pair interactions of the single-site CG OTP at three state points: (a) 380 K, 1.029 g/ml, (b) 600 K, 1.025 g/ml, (c) 700 K, 1.069 g/ml. Due to the free energy nature of CG interactions, the OTP CG interactions vary with temperature and density, as evident when compared to panel (a) that is represented by green dots.}
\end{figure}
\begin{figure}
\includegraphics[width=0.96\columnwidth]{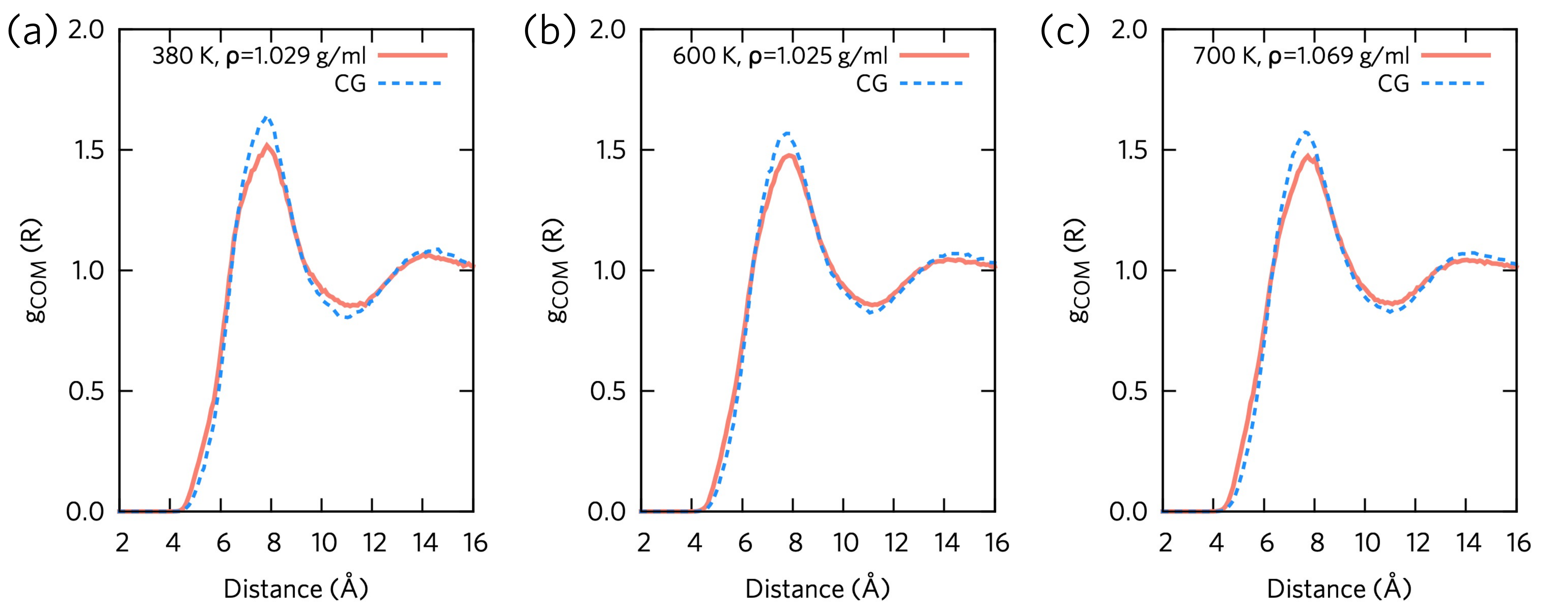}
\caption{\label{fig:OTP-RDF} Center-of-mass pair correlation functions $g_\mathrm{COM}(R)$ for atomistic (red lines) and the single-site CG (blue dots) simulations of OTP at three state points: (a) 380 K, 1.029 g/ml, (b) 600 K, 1.025 g/ml, (c) 700 K, 1.069 g/ml. Over a wide range of temperatures ($\Delta T = 320$ K) and densities ($\Delta V=11561\,\angstrom^3$), the spatial CG approach reproduces structural correlations.}
\end{figure}
In Fig. \ref{fig:OTP-interaction}, the parametrized CG interactions are illustrated for three different temperature and density conditions out of 23 state points. For clarity, the CG interactions at the remaining state points are provided in the SM (Sec. III). The CG OTP interactions are consistently repulsive, regardless of the thermodynamic conditions. Generally, $U(R)$ is always positive and appears to decay around 14--16 $\angstrom$ with subtle variations in positions and slopes. These changes arise from the free energy nature of bottom-up CG interactions, as expressed in Eq. (\ref{eq:CGPMF}); $U_\textrm{CG}$ varies with temperature, pressure, and other thermodynamic state variables, as extensively demonstrated in liquid systems \cite{allen2009evaluating,lu2011multiscale,hsu2015thermomechanically,jin2019understanding,jin2020temperature}.

The purely repulsive profile signifies the strongly correlating nature of CG OTP, akin to an IPL form. Nevertheless, as these bottom-up CG interactions are state point-dependent, before delving into the density scaling of CG OTP, it is imperative to validate whether the resulting CG interactions can faithfully reproduce the important structural correlations observed at the microscopic level.

\subsubsection{Validating CG Models: Structural Correlations}
To gauge the performance of the CG models of OTP, we perform CG simulations utilizing the obtained interactions and compute the radial distribution function (RDF) of the center-of-mass configurations, i.e., the intermolecular RDF. For each thermodynamic state point, we constructed a separate CG system and conducted \textit{NVT} dynamics for 5 ns, employing the same thermostats as in the atomistic simulation. Using the last snapshot from the atomistic simulation, we performed center-of-mass coarse-graining to generate the initial configuration for the CG runs. The CG RDFs are then calculated from the CG simulations (Fig. \ref{fig:OTP-RDF}), and the CG RDFs at other thermodynamic state points are presented in the SM Fig. S3.

In spite of the purely repulsive nature of the interaction profile, our observation indicates that the CG simulation can faithfully reproduce the atomistic RDF across all thermodynamic state points studied. Even though high-resolution structural characteristics are inevitably lost at this level of coarse-graining, the RDFs exhibit subtle structuring near 4-6 $\angstrom$ and the first peak around 8 $\angstrom$. Notably, the CG models capture the general shape of $g(R)$ with a maximum difference of 0.1. Despite the approximations introduced such as pairwise decomposability, we conclude that the CG models capture the essential structural correlations of OTPs. 

The purely repulsive CG interaction may seem inconsistent with the commonly adopted potential for the three-site Lewis-Wahnstr\"{o}m model comprising non-bonded interactions of Lennard-Jones form. In Appendix B, we demonstrate that the bottom-up CG interaction is, nevertheless, consistent with this phenomenological model, suggesting room for improvement of liquid-state model development.

\subsection{Density Scaling of OTP}
\subsubsection{Scaling Exponent} \label{subsec:representability}
Having established that the spatial CG OTP models are representative of microscopic correlations in a renormalized manner, our focus now shifts to understanding the density scaling relationship of OTP through spatial coarse-graining. Similar to temporal coarse-graining, at the single-site CG resolution, we anticipate that the vibrational motions resulting from the intramolecular interactions are integrated out by the coarse-graining process. In Appendix C, we validate this hypothesis by comparing the power spectrum from CG MD simulations with that of the atomistic reference. The comparison reveals that high-frequency vibrations are effectively integrated out at the CG resolution, leaving only translational motions. Along with the purely repulsive profile of OTP interaction, this analysis demonstrates that spatial coarse-graining is a highly effective strategy for establishing density scaling of OTP from first principles.

We emphasize that the conventional approaches for isomorph theory are limited in their application to (spatial) CG systems. This limitation arises from the renormalized degrees of freedom during the coarse-graining process, causing the thermodynamic properties of CG models to deviate significantly from the atomistic reference. This issue, known as the \textit{representability issue} in CG modeling \cite{johnson2007representability,dunn2016van,wagner2016representability}, implies that one cannot use the na\"ively evaluated thermodynamic properties, as at the atomistic level, please refer to the discussion in Appendix {D}. In this light, we propose a direct estimation of $\gamma_s$ for spatial CG models using the formal definition of $\gamma$ at any point in the thermodynamic phase diagram, 
\begin{equation} \label{eq:CGGamma}
    \gamma_s = \left( \frac{\partial \ln T}{\partial \ln \rho}\right)_{S_\textrm{ex}},
\end{equation}
where we distinguish the exponents from temporal and spatial CG approaches as $\gamma_t$ and $\gamma_s$, respectively. Equation (\ref{eq:CGGamma}) implies that $\gamma_s$ is the (logarithmic) slope of the isomorph (configurational adiabat) through the state point $(\rho, T)$. While evaluating $\gamma_s$ strictly using Eq. (\ref{eq:CGGamma}) is less common, we will show that this method provides a viable way to estimate $\gamma_s$ for spatial CG models. Yet, it should be noted that Eq. ($\ref{eq:CGGamma}$) needs to be evaluated at the CG level, requiring an estimation of the excess entropy of CG models beforehand, which we now discuss in Subsection C 2. 

\subsubsection{Estimation of Excess Entropy in CG Systems}
In order to estimate the excess entropy ($S_\textrm{ex}$) based on statistical mechanical theories, Wallace \cite{wallace1987correlation} as well as Baranyai and Evans \cite{baranyai1989direct}
showed that $S_\textrm{ex}$ can be expressed as a systematic expansion over $n$-particle distribution functions \cite{green1965molecular}, 
\begin{equation} \label{eq:Wallace}
    S_\textrm{ex} = \sum_{n\ge2} S^{(n)}_\textrm{ex},
\end{equation}
where $S^{(n)}_\textrm{ex}$ is the \textit{n}-particle contribution to the excess entropy. While for simple liquids, the two-body contribution is often dominant and enough to estimate the excess entropy, 
\begin{align} \label{eq:2bdsex}
    S_\textrm{ex}&\approx S_\textrm{ex}^{(2)} \nonumber \\ &= -2\pi \rho \int_0^\infty \mathrm{d}R\left\{g(R)\ln g(R)-[g(R)-1] \right\}R^2,
\end{align}
and this approach to OTP is questionable since many liquids, e.g., water \cite{kumar2009tetrahedral,cisneros2016modeling},  exhibit higher-order contributions beyond the pairwise level due to orientational and many-body correlations \cite{jin2023hierarchical}. 

More importantly, as depicted in Fig. \ref{fig:OTP-RDF}, the RDF of OTPs remains remarkably similar across a broad range of temperature and density conditions. This suggests that changes in thermodynamic state variables minimally impact the two-body correlation level, whereas complexities may arise beyond pairwise correlations. Unfortunately, the assessment of higher-order orientational correlations requires extensive sampling and approximations \cite{lazaridis1996orientational}, making it impractical to compute higher-order contributions for more complex molecules \cite{lazaridis2000solvent,zielkiewicz2008two}.

\subsubsection{Modal Entropy of CG OTP}
We adopt an alternative approach based on the two-phase thermodynamics (2PT) method \cite{lin2003two,lin2010two,pascal2011thermodynamics, jin2015mechanisms}, which has recently demonstrated its utility in computing the excess entropy of molecular CG liquids. The fundamental idea of this method originates from the quasiharmonic analysis by Karplus and coworkers \cite{karplus1981method}, later formulated by Goddard et al. \cite{lin2003two,lin2010two}. The original formulation of the 2PT method was derived at the fully atomistic level, and thermodynamic properties can be estimated by constructing the partition function $\mathcal{Q}$ from the density of state (DoS) of the liquid, denoted as $D(\nu)$, with the appropriate weighting function $W(\nu)$, 
\begin{eqnarray} \label{eq:lnQ}
    \ln \mathcal{Q} = \int_0^\infty D(\nu)W(\nu) d\nu.
 \end{eqnarray}
The $D(\nu)$ can be directly obtained from the microscopic MD simulations by taking the Fourier transformation of the velocity autocorrelation functions. However, estimating various thermodynamic properties using Eq. (\ref{eq:lnQ}) and assuming that the system follows a quantum harmonic oscillator results in singularities at $\nu=0$. The 2PT method circumvents this issue by decomposing the liquid DoS into a combination of two phases, namely gas-like and liquid-like phases: $D_\mathrm{liq}(\nu) = f\times D_\mathrm{gas}(\nu)+(1-f)D_\mathrm{solid}(\nu)$. Here, $f$ is the ``fluidicity'' factor introduced to account for diffusive contributions. This decomposition yields the well-defined zero-frequency DOS as $D_\textrm{gas}(0)=D(0)$, which can be computed from the Carnahan-Starling equation of state \cite{carnahan1970thermodynamic}, and $D_\textrm{solid}(0)=0$, allowing for the estimation of thermodynamic properties. The fluidicity factor $f$ is determined by the ratio between the self-diffusivity of the system and the hard sphere diffusivity using the Chapman-Enskog theory \cite{chapman1990mathematical}. 

The 2PT method provides an efficient and accurate way of estimating entropy through $S = k_B T \frac{\partial \ln \mathcal{Q}}{\partial T}+k_B \ln \mathcal{Q}$. Furthermore, various other thermodynamic properties, such as heat capacity or Helmholtz energy, can be calculated from the constructed DoS in a similar manner. Despite its approximate nature, it has been demonstrated that the thermodynamic properties predicted for common organic liquids using the 2PT approach are in close agreement with experimental results and more rigorous perturbation methods \cite{pascal2011thermodynamics}. For OTP, we further substantiate the consistent performance by calculating the standard molar entropy at ambient temperatures: $S=252.67$ J/mol/K (260 K) and $350.54$ J/mol/K (360 K). These values reasonably agree with experimental observations from calorimetric data, with a relative error of approximately 10 \% \cite{chang1972heat}. This agreement implies that the 2PT method can be employed as a robust and efficient approach for estimating excess entropy to understand the density scaling of CG OTP. 

A significant advantage of the 2PT method for computing the excess entropy of CG systems lies in its capability to decompose entropy into different modes. This is feasible by decomposing atomic velocities into translational, rotational, and vibrational contributions for each molecule \textit{i}: $v = v^\mathrm{trn}(i)+v^\mathrm{rot}(i)+v^\mathrm{vib}(i)$. The translational velocity $v^\mathrm{trn}(i)$ is understood as the center-of-mass velocity, while the rotational velocity $v^\mathrm{rot}(i)$ is estimated by treating the system as a rigid rotor, i.e., $v^\mathrm{rot}(i) = \omega(i)\times v^\mathrm{tot}(i)$. The angular velocity $\omega(i)$ is estimated by inverting $\mathbf{L}(i)=\mathbf{I}_i \omega (i)$, where $\mathbf{L}(i)$ is the angular momentum of molecule $i$, $\mathbf{L}(i)=\sum_j m_j (\mathbf{r}_j \times \mathbf{v}_j)$ with atom $j$ in $i$, and $\mathbf{I}_i$ is the inertia tensor. Finally, the vibrational velocity can be computed as the complement, i.e., $v^\mathrm{vib}(i)=v^\mathrm{tot}-\left(v^\mathrm{trn}(i)+v^\mathrm{rot}(i) \right)$.

From the decomposed modal velocities, one can construct the modal DoS and the corresponding entropy using the appropriate weighting functions. Detailed discussions and formulas for the weighting functions are given in the original 2PT literature \cite{lin2003two,lin2010two,pascal2011thermodynamics, jin2015mechanisms} and extensively discussed in recent applications to CG models \cite{jin2019understanding,jin2023understanding}. In summary, the 2PT approach enables efficient estimation and decomposition of entropy as $S_\mathrm{FG} = S_\mathrm{FG}^\mathrm{trn}+S_\mathrm{FG}^\mathrm{rot}+S_\mathrm{FG}^\mathrm{vib}.$

\subsubsection{Excess Entropy of CG OTP}
The modal decomposition of entropy provides a practical starting point for estimating excess entropy, particularly in CG modeling. The underlying reasoning is twofold. First, when calculating the ideal gas entropy of the target system to assess the excess entropy, this decomposition allows for specifying the modal contribution to an ideal gas at the given CG resolution \cite{mcquarrie1997physical}. This advantage is more pronounced at the single-site CG resolution chosen in this study \cite{jin2019understanding}. At this single-site level, where there are no rotational and vibrational motions, the CG entropy arises entirely from translational motions. In this case, the ideal gas entropy can be directly estimated from the Sackur-Tetrode equation:
\begin{align} \label{eq:trn}
    s_{id}^\mathrm{trn} = \frac{S_{id}^\mathrm{trn}}{Nk_B} = - \ln \left( \frac{h^2 }{2\pi m k_B T}\right)^\frac{3}{2} -\ln \left( \frac{N}{V} \right)+\frac{5}{2},
\end{align}
and then the CG excess entropy is estimated from the translational contribution,
\begin{align} \label{eq:sex_trn}
s_{ex}^\mathrm{CG}&=s_{ex}^\mathrm{trn} = s^\mathrm{CG} - s_{id}^\mathrm{trn} \nonumber \\ &= s^\mathrm{CG}- \left[\frac{5}{2} - \ln \left( \frac{h^2 }{2\pi m k_B T}\right)^\frac{3}{2} -\ln \left( \frac{N}{V} \right)\right].
\end{align}
The second advantage of the single-site CG mapping lies in the useful feature of directly deriving the missing entropy from the coarse-graining procedure, i.e., orientational (rotation + vibration) entropy, as only the translational degrees of freedom remain at the CG level. Therefore, under the limit of perfect spatial coarse-graining, one can determine the difference of the FG and CG entropy as $s^\mathrm{FG}-s^\mathrm{CG}=s^\mathrm{rot}+s^\mathrm{vib}$ \cite{jin2019understanding}.

At the CG resolution, Ref. \onlinecite{jin2019understanding} numerically confirmed that the difference between the FG and CG entropy values corresponds to the contribution from the missing (intramolecular) degrees of freedom. Since our primary focus in this section is to assess the isomorph of the spatial CG OTP model, we employ Eq. (\ref{eq:sex_trn}) for the conditions studied earlier. For each CG model, we calculate the translational entropy using $v^\mathrm{trn}$ with the 2PT method, and the corresponding ideal gas entropy is then estimated by the molecular weight of OTP and number density (given as the $N/V$ term). We note that the entropic contribution from intramolecular degrees of freedom can be estimated by utilizing Eqs. (\ref{eq:rot}) and (\ref{eq:vib}). Furthermore, by including the contribution of $S_\textrm{FG}^\textrm{rot}$ in addition to $S_\textrm{FG}^\textrm{trn}$, the performance of the approximate Lewis-Wahnstr{\"o}m model can be directly assessed by the entropy values. A detailed analysis is provided in Appendix E.

\subsubsection{Results and Summary}
Figure \ref{fig:OTP-gamma} illustrates the changes in the estimated excess entropy in the $(\rho, T)$ phase diagram across a wide range of simulated thermodynamic state points. Within these state points, the (dimensionless) excess entropy values range from $-8.5$ to $-4.5$. Generally, we observe that as temperature increases at constant density, the excess entropy decreases as the system becomes more ideal gas-like. Likewise, at a fixed temperature, the excess entropy decreases as the density decreases (resulting in a longer box length). These trends are consistent with expectations, and, based on the computed $s_\textrm{ex}$ from our approach over the selected thermodynamic state points, we interpolated the contour along this phase diagram corresponding to the configurational adiabats, i.e., $s_\textrm{ex}=\textrm{(constant)}$. 

While there are some fluctuations in higher-density regions, the configurational adiabats at relatively higher temperatures exhibit similar slopes. As observed from the temporal coarse-graining approach, spatial coarse-graining also exhibits a $\gamma_s$ that depends on density and other thermodynamic state variables. Since our primary target is to derive the density scaling exponent from the estimated $\gamma_s$ values and considering the approximations and interpolations in Fig. \ref{fig:OTP-gamma}, we are mainly interested in the general behavior of $\gamma_s$ in CG models of OTP near these regions of the phase diagram. This is akin to determining the average $\gamma_t$ from coarse-graining in time, which we found as $\gamma_t=6.3$. In spatial coarse-graining, averaging $\gamma$ values along the isomorph yields $\gamma_s=6.8$, showing a reasonably good agreement with $\gamma_t$. The difference between $\gamma_s$ and $\gamma_t$ can be understood in terms of the nature of coarse-graining in time, where separating fast relaxation from slow relaxation motions is often challenging due to the complexity of these motions. In other words, even at the optimal $\tau_\alpha$, the remaining characteristic of fast relaxation might lead to slightly lower $\gamma$ values and correlations ($R$) across the three different approaches. In spatial coarse-graining, the fast degrees of freedom are fully integrated out, however, and only slowly relaxing non-bonded interactions remain at the CG level, resulting in a slightly larger $\gamma$.

Finally, the fidelity of our coarse-graining approach is assessed by comparing it to the experimental scaling relationship. Since pure OTP undergoes slow crystallization in typical experimental settings \cite{Takahara1999}, we compared our $\gamma$ values to experimental data obtained from OTP/ortho-phenylphenol (OPP) mixtures that closely resemble our system. In Refs. \onlinecite{Roland2004, Roland2005}, the experimental $\gamma$ was measured as 6.2, which is in quantitative agreement with our findings. This remarkable consistency further corroborates the validity of our results and the fidelity of our CG approach.

\begin{figure}
\includegraphics[width=0.75\columnwidth]{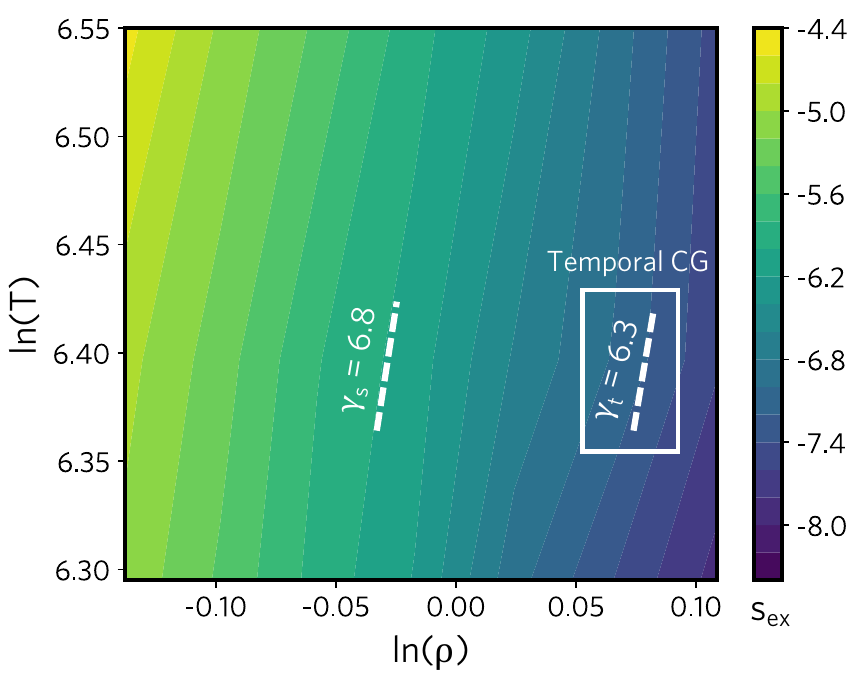}
\caption{\label{fig:OTP-gamma} Excess entropy and density scaling exponent estimated from the spatial coarse-graining approach by numerically estimating the $\partial \ln T/\partial \ln \rho$ along curves of constant $S_\textrm{ex}$ (units: $T$ in K and $\rho$ in g/ml). The single-site CG excess entropy was estimated across various thermodynamic state points using Eq. (\ref{eq:sex_trn}). Averaging over the illustrated state points gives $\gamma_s=6.8$. The value $\gamma_t = 6.3$ from temporal coarse-graining is shown as a reference (right dashed line).}
\end{figure}

\section{Density Scaling of OTP Dynamics from First Principles}\label{sec:Outlook}
\subsection{Microscopic Theory of Density Scaling}
The two distinct coarse-graining approaches in time and space enable the \textit{ab initio} estimation of the density scaling exponent from microscopic (atomistic) simulations, as described by Eq. (\ref{eq:densityscaling}). However, as previously noted, the microscopically determined $\gamma$ and Eq. (\ref{eq:densityscaling}) do not specify the exact form of the scaling relationship. We now derive a microscopic theory of density scaling by introducing excess entropy scaling in liquids to link $S_\mathrm{ex}$ to $\rho^\gamma/T$. Inspired by the Avramov model \cite{avramov2000pressure}, the excess entropy $S_\mathrm{ex}$ can be expressed as a function of the heat capacity $c_v$, the temperature $T$, and the specific volume $\nu\equiv\rho^{-1}$,
\begin{equation} \label{eq:avramov_entropy}
    S_\mathrm{ex} = c_v \ln \left( \frac{T\nu^\gamma}{T_\mathrm{ref}\nu_\mathrm{ref}^\gamma} \right).
\end{equation}
Here, a reference state (``ref'') is chosen to approximate ideal gas behavior with $S_\mathrm{ref}\approx S_\mathrm{id}$, and we further assume that $c_v$ and $c_p-c_v$ do not change significantly with respect to $T$ and $V$, respectively. For a detailed derivation, we refer to Appendix F. Note that Eq. (\ref{eq:avramov_entropy}) ensures a physically realizable excess entropy ($S_\mathrm{ex}<0$) with $T\nu^\gamma < T_\mathrm{ref}\nu_\mathrm{ref}^\gamma$. Substituting Eq. (\ref{eq:avramov_entropy}) into excess entropy scaling yields
\begin{equation}
    \tilde{D} = D_0 \exp \left[ \frac{\alpha c_v}{Nk_B} \ln \left(\frac{T\nu^\gamma}{T_\mathrm{ref}\nu_\mathrm{ref}^\gamma} \right)\right],
\end{equation}
implying that density scaling can be derived as
\begin{eqnarray} \label{eq:densityscaling1}
    \ln\left[\frac{1}{\tilde{D}}\right] = \ln\left[\frac{1}{D_0} \cdot \left(\frac{T_\mathrm{ref}}{\rho_\mathrm{ref}^\gamma}  \right)^{\frac{\alpha c_v}{Nk_B}} \right]+ \frac{\alpha c_v}{Nk_B} \ln\left( \frac{\rho^\gamma}{T} \right).
\end{eqnarray}

If $\rho^\gamma/T$ is not too far from the reference value $\rho_\mathrm{ref}^\gamma/T_\mathrm{ref}$, Eq. (\ref{eq:densityscaling1}) can be further approximated as
\begin{equation} \label{eq:densityscaling2}
    \ln\left[\frac{1}{\tilde{D}}\right] \approx  \frac{\alpha c_v}{Nk_B}\cdot \frac{T_\mathrm{ref}}{\rho_\mathrm{ref}^\gamma} \left( \frac{\rho^\gamma}{T} \right) + \left( \ln\left[\frac{1}{D_0} \right]- \frac{\alpha c_v}{Nk_B}\right),
\end{equation}
which indicates that the relationship between $\ln(\tilde{D}^{-1})$ and $\rho^\gamma/T$ is approximately linear with slope $(\alpha c_v T_\mathrm{ref})/(Nk_B \rho_\mathrm{ref}^\gamma)$. 
Equations (\ref{eq:densityscaling1}) and (\ref{eq:densityscaling2}) are the central results of this section, because they provide microscopic origins underlying density scaling through the introduction of excess entropy scaling.

\subsection{Prediction of Density Scaling}
Another key advantage of our computational framework is its capability to predict whether a complex system, i.e., realistic molecular liquids, will exhibit density scaling (or hidden scale invariance), and if so, to determine the scaling exponent. This is particularly crucial for bridging the great divide between experiments, where exploring the low-temperature glassy state is often challenging, and theory, which has largely concentrated on simplified, non-realistic model systems.

\begin{figure}
\includegraphics[width=0.8\columnwidth]{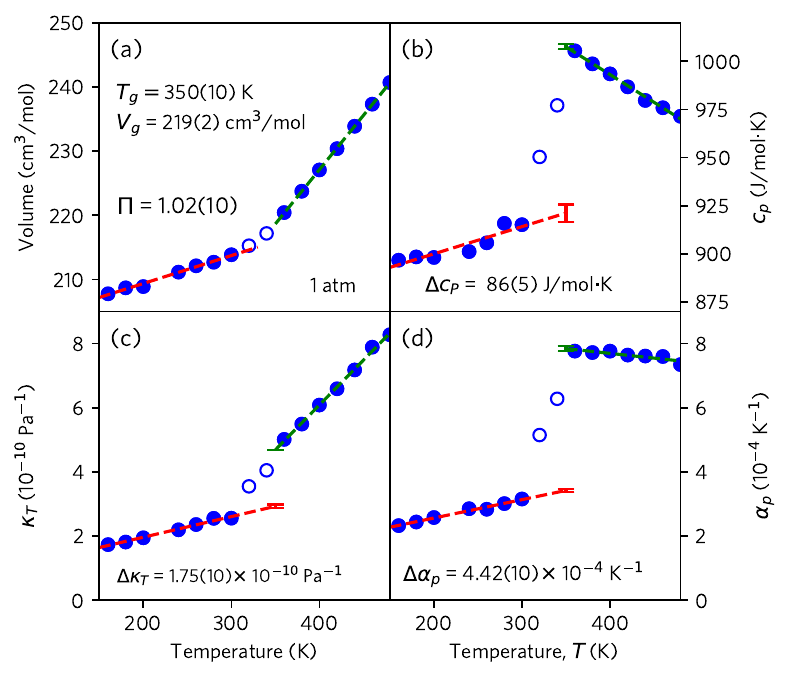}
\caption{\label{fig:prigogine_defay} Microscopic-level prediction of density scaling of OTP by computing the Prigogine-Defay ratio $\Pi=1.02(10)$ from fully atomistic simulations at the glass transition temperature $T_g=350(10)$ K using the jumps of the following response functions between the red and green dashed lines: (a) Average volume in constant $NpT$ simulations, (b) specific isobaric heat capacity, $c_p$, (c) isothermal compressibility, $\kappa_T$, (d) thermal expansion coefficient, $\alpha_p$.}
\end{figure}

Based on the microscopic computer simulation, we can validate density scaling by estimating the classical Prigogine-Defay ratio, $\Pi$ [Eq.\ (\ref{eq:prigogine_classic})]. As discussed in the Introduction, strongly correlating systems, i.e., systems exhibiting an effective single-parameter phase diagram, typically obey $\Pi \approx 1$ \cite{chemical_thermodynamics, Gupta1976, Takahara1999, Schmelzer2006, Ellegaard2007, Pedersen2008b, Gundermann2011, Fragiadakis2011, Casalini2011, Tropin2012, Garden2012}. Therefore, estimating $\Pi$ for OTP from first principles provides an indirect prediction of density scaling. Using the computational details outlined in Appendix G, we calculated the $\Pi$ value near the glass transition and found $\Pi=1.02(10)$ (Fig.\ \ref{fig:prigogine_defay}), which strongly indicates the existence of density scaling in OTP. Moreover, the estimated thermodynamic properties $c_P$, $\kappa_T$, $\alpha_p$ from this analysis can be used to estimate the slopes derived from Eqs. (\ref{eq:densityscaling1}) and (\ref{eq:densityscaling2}) (see Subsection D). 

At the CG level, the existence of density scaling can be directly predicted by the consistency of the estimated exponent $\gamma$ across CG models. Specifically, if the system exhibits density scaling, we expect both spatial and temporal exponents from coarse-graining approaches to yield similar values across different thermodynamic state points. The consistency observed in OTP supports the existence of a one-dimensional phase diagram.

\subsection{FG Excess Entropy Scaling of OTP}
Having demonstrated that OTP exhibits density scaling, we first evaluate $\alpha$  in Eqs. (\ref{eq:densityscaling1}) and (\ref{eq:densityscaling2}) by employing excess entropy scaling at the fully atomistic level. Unlike at the single-site CG level, it is crucial to account for the two ``missing'' orientational contributions, $s_\textrm{ex}^\mathrm{rot}$ and $s_\textrm{ex}^\mathrm{vib}$, in addition to the translational component. While these orientational components are particularly challenging to estimate for the relatively complex OTP system, we can evaluate them by \cite{jin2023understanding}
\begin{align} \label{eq:rot}
    s_{id}^\mathrm{rot} = \ln \left[ \frac{\sqrt{\pi}}{\sigma}\left( \frac{T^3 e^3}{\Theta_A \Theta_B \Theta_C}\right)^\frac{1}{2}\right],
\end{align}
and
\begin{align} \label{eq:vib}
    s_{id}^\mathrm{vib} = \sum_{j=1}^{3N-6} \left[ \frac{\Theta_{v_j}/T}{e^{\Theta_{v_j}/T}-1} - \ln \left(1-e^{-\Theta_{v_j}/T} \right)\right].
\end{align}
In Eq. (\ref{eq:rot}), $\Theta_{A,B,C}$ represents the characteristic rotation temperatures along the $x, y,$ and $z$ axes with the rotational symmetry number $\sigma$, while $\Theta_{v_j}$ in Eq. (\ref{eq:vib}) is the characteristic vibrational temperature at the $j$-th vibrational mode $v_j$. 
Altogether, the excess entropy of atomistic systems can be estimated as
\begin{align} \label{eq:sex_fg}
    s_\textrm{ex}^\mathrm{FG} & = s_\textrm{ex}^\mathrm{trn}+s_\textrm{ex}^\mathrm{rot}+s_\textrm{ex}^\mathrm{vib} \nonumber \\ 
                       & = s^{FG} - \left[\frac{5}{2} - \ln \left( \frac{h^2 }{2\pi m k_B T}\right)^\frac{3}{2} -\ln \left( \frac{N}{V} \right)\right] \nonumber \\ &-\ln \left[ \frac{\sqrt{\pi}}{\sigma}\left( \frac{T^3 e^3}{\Theta_A \Theta_B \Theta_C}\right)^\frac{1}{2}\right] \nonumber \\
                       & - \sum_{j=1}^{3N-6} \left[ \frac{\Theta_{v_j}/T}{e^{\Theta_{v_j}/T}-1} - \ln \left(1-e^{-\Theta_{v_j}/T}\right)\right].
\end{align}
Reference \onlinecite{pascal2011thermodynamics} demonstrated that Eq. (\ref{eq:sex_fg}) yields almost identical entropy values for simple liquids compared to those calculated by computationally expensive thermodynamic integration methods. Relatedly, Ref. \onlinecite{jin2023understanding} recently demonstrated that Eq. (\ref{eq:sex_fg}) can accurately compute the excess entropy of complex molecules beyond the pairwise contribution inferred from the RDF. 

\begin{figure}
\includegraphics[width=0.8\columnwidth]{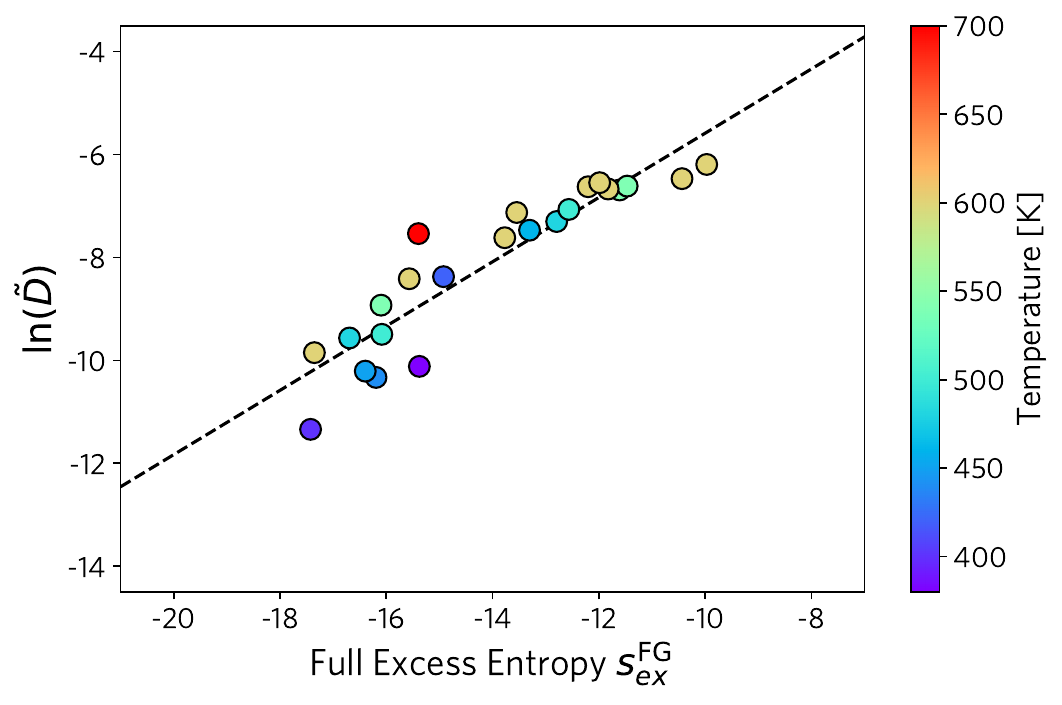}
\caption{\label{fig:sex_scaling_fg} Excess entropy scaling of OTP from atomistic simulation using Eq. (\ref{eq:sex_fg}) with $\alpha\approx 0.61$ (dashed lines) colored based on the temperature.}
\end{figure}
In practice, in contrast to simple liquids, where estimating $s_{id}^\mathrm{rot}$ and $s_{id}^\mathrm{vib}$ is relatively straightforward using Eqs. (\ref{eq:rot}) and (\ref{eq:vib}), or state-of-the-art equations of state like {\tt{REFPROP}} \cite{lemmon2002nist}, the complexity of OTP requires accounting for $90$ degrees of freedom for vibrations and three for rotations, which is computationally prohibitive to directly estimate from MD simulations. To address this challenge, we infer the characteristic temperatures for Eq. (\ref{eq:sex_fg}) using quantum chemical spectroscopy calculations \cite{otp-nist}, as detailed in the SM (Sec. IV). The final scaling relationship depicted in Fig. \ref{fig:sex_scaling_fg} for OTP demonstrates that the atomistic OTP follows excess entropy scaling with $\alpha\approx0.61$ 
\begin{equation}\label{excessscaling}
\tilde{D} = D_0 \exp (0.61 s_\mathrm{ex}^\mathrm{FG}).
\end{equation}

The substantial contributions of $s_{ex}^\mathrm{rot}$ and $s_{ex}^\mathrm{vib}$ to $s_{ex}^\mathrm{FG}$ underscore the importance of considering configurational entropy beyond simple pair entropy. While deviations seen in Fig. \ref{fig:sex_scaling_fg} might arise from the characteristic temperatures used, which were not directly derived from MD simulation, our findings mark the first attempt to demonstrate full excess entropy scaling from atomistic OTP simulations.

\subsection{First-Principles Density Scaling}
As a final analysis, we now examine whether our microscopic theory can provide a correct density scaling relationship using the exponent $\gamma$ determined by the proposed CG approaches. While the experimental success of density scaling for OTP motivates this line of investigation \cite{Tolle1998, Tolle2001}, it remains uncertain whether molecular-level simulations would exhibit this scaling \textit{a priori}, given that no demonstration of density scaling from full atomistic simulations has been reported.

\begin{figure}
\includegraphics[width=0.75\columnwidth]{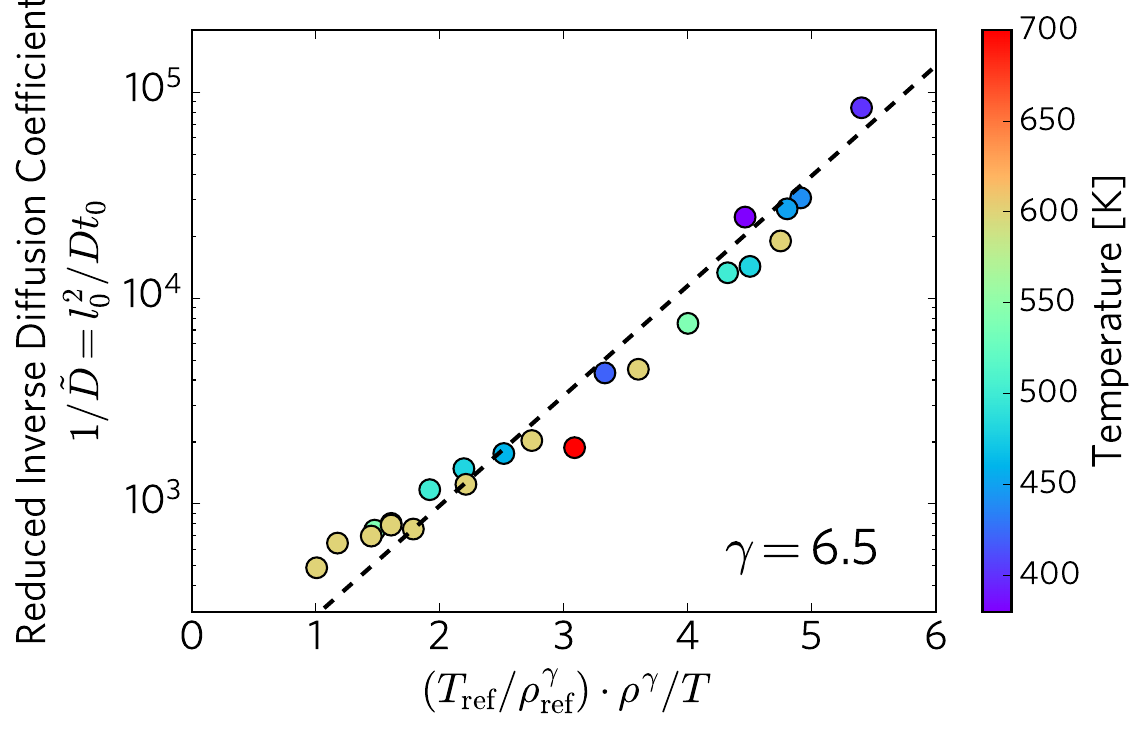}
\caption{\label{fig:density_scaling_otp} Density scaling of OTP from atomistic simulation, where the dynamical properties from fully atomistic simulations are scaled using the scaling exponent $\gamma=6.5$ determined from both coarse-graining in space and in time over a wide range of temperatures and densities.}
\end{figure}

Density scaling was performed on atomistic OTP simulations by scaling the computed diffusion coefficients (see SM Sec. I) into the inverse reduced diffusion coefficient $1/\tilde D = l_0^2/Dt_0$ (where $l_0=\rho^{-1/3}$ and $t_0=l_0\sqrt{m/k_BT}$, thus $\tilde{D}$ is dimensionless). Subsequently, we estimated the remaining prefactor in Eq. (\ref{eq:densityscaling2}). First, we approximated the reference state as the boiling point, identified as $T=600$ K and $\rho=0.877$ g/ml (see Appendix G), yielding a constant factor of $T_\mathrm{ref}/\rho_\mathrm{ref}^\gamma= 1400$ (g/mL)$^\gamma/\mathrm{K}$. Over a broad range of temperatures and densities up to the reference state, we confirmed that $c_v$ fluctuates mildly around $895$ J/mol$\cdot$K, which is consistent with our approximation. Combining these results, our microscopic theory imparts the linear scaling relationship of
\begin{equation}\label{eq:densityscalingfinal}
    \ln\left[\frac{1}{\tilde{D}}\right] \approx  0.532\cdot\left( \frac{1400\rho^\gamma}{T} \right) + \mathrm{(constant)},
\end{equation}
where the slope 0.532 is estimated from $\alpha c_v/(Nk_B)$, and the constant term includes the remainder from Eq. (\ref{eq:densityscaling2}) along with higher-order expansion terms. To validate Eq. (\ref{eq:densityscalingfinal}), Fig. \ref{fig:density_scaling_otp} illustrates density scaling of $1/\tilde D$ against $1400\,\rho^\gamma/T$, where we used the average density scaling exponent $\gamma=6.5$ derived from both coarse-graining in space and time. Remarkably, in Fig. \ref{fig:density_scaling_otp}, we observe that the values of the inverse diffusion coefficients collapse onto the linear curve predicted by Eq. (\ref{eq:densityscalingfinal}).

The main messages of Fig. \ref{fig:density_scaling_otp} are \textit{twofold}. First, Fig. \ref{fig:density_scaling_otp} demonstrates that our CG approach faithfully captures the density scaling relationship with the estimated exponent $\gamma$, consistent with experimental results. Second, and more critically, our microscopic theory successfully generates a master curve and enables the prediction of diffusion coefficients beyond the conditions studied, particularly at lower temperatures where dynamics slows down significantly, making experimental and atomistic simulations impractical \cite{berthier2023modern}. Altogether, this approach represents the first systematic demonstration of density scaling at the molecular level and a significant leap forward in the pursuit of bottom-up approaches for complex systems that are experimentally challenging and beyond the reach of atomistic simulations.

\section{Conclusion}
In the exploration of soft condensed matter, density scaling emerges as an invaluable framework for understanding how transport properties, such as viscosity and diffusion coefficients, scale with changes in density and temperature. While numerous experiments have evidenced this relationship in realistic liquids, the deeper microscopic nature of this scaling has remained elusive. The concept of hidden scale invariance in the isomorph theory offers physical insights to address this gap, but these theories are traditionally limited to point particle descriptions. Extending these concepts to molecules presents challenges that have limited their broader application to more complex systems, e.g., biomolecules and polymers.

This work establishes the first bottom-up molecular-level correspondence to density scaling for realistic materials, extending and enhancing our understanding beyond existing semiempirical studies on simplified molecular liquids \cite{koperwas2019effect,Roland2019,kaskosz2023origin,kaskosz2024breakdown}. To achieve this, we introduce two bottom-up coarse-graining approaches applied to temporal and spatial domains to efficiently explore the free energy landscape of atomistic OTP. In the temporal coarse-graining approach, we time-average the instantaneous fluctuations of potential energy, allowing us to assess time-dependent slow fluctuations of OTP molecules and obtain the frequency-dependent scaling exponent from the fluctuation-dissipation theorem. For the spatial coarse-graining approach, the less important intramolecular degrees of freedom are renormalized by reducing the OTP molecule to a single-site center-of-mass representation. The spatial CG scaling coefficient is then calculated by tracing out the isomorphs through the estimation of the CG excess entropy. While these two bottom-up coarse-graining approaches stem from different assumptions and microscopic principles, we find that the density scaling coefficients obtained from these distinct approaches exhibit quantitatively similar values, consistent with experimental observations, when the system shows hidden scale invariance.

Next, by utilizing the microscopically derived $\gamma$ to perform density scaling for atomistic diffusion coefficients, we establish that the estimated $\gamma$ provides the correct scaling across a broad range of temperatures and densities. This approach facilitates efficient computation of phase diagrams within a single-dimensional $\rho^\gamma/T$ space.

Finally, by integrating excess entropy scaling with density scaling, we derive a first-principles theory of density scaling, providing a microscopic handle for predicting transport coefficients in regimes that are beyond the reach of simulation, even for simple model systems like Lennard-Jones. As understanding long-time scale dynamical properties at the microscopic level is crucial for advancing predictable multiscale modeling, we expect that our CG framework will be broadly applicable to a wide range of molecular liquids.

In summary, our research program imparts valuable insights into \textit{whether a liquid exhibits hidden scale invariance and, if that is the case, how the corresponding exponents and scaling behaviors can be predicted}. Our theory and its computational implementation are highly general and open-sourced \cite{zenodo}, making them readily applicable to other molecular liquids. Our approach can also be combined with experimental data to significantly enhance the understanding of density scaling. Building on the developments presented here, this work paves the way toward solving key remaining questions surrounding hidden scale invariance in condensed matter. These include uncovering the fundamental reasons why some systems exhibit density scaling while others do not, and establishing a rigorous correspondence between CG approaches in space and time. 
By leveraging our bottom-up coarse-graining technology and the concept of density scaling, our work opens new avenues for microscopically informed explorations of dynamics in complex molecules.

\section*{Data Availability}
The data that support the findings of this study are available within the Supplemental Material \cite{supp} and openly available in the {\tt{Zenodo}} repository at http://doi.org/10.5281/zenodo.11624467 \cite{zenodo}.

\begin{acknowledgments}
J.J. thanks the Arnold O. Beckman Postdoctoral Fellowship for funding and academic support. The authors also appreciate the valuable comments and helpful feedback during the ``Viscous liquids and the glass transition (XIX)'' workshop in 2023. This work was supported by the VILLUM Foundation's \textit{Matter} grant (VIL16515).
\end{acknowledgments}

\section*{Appendices}
\subsection*{A. Spatial Coarse-Graining: Force-matching}
\subsubsection{Numerical Settings}
The construction of the spatial CG model for OTP proceeded as follows. From atomistic MD trajectories, we mapped OTP molecules to their center-of-masses. The mapping operator $M_\mathbf{R}^N$ maps the configuration of each molecule to its center-of-mass with the accumulated force $\mathbf{f}_I(\mathbf{r}^n)=\sum_{i\in \mathcal{I}_I}\mathbf{f}_i^\mathrm{FG}$, where $\mathbf{f}_i^\mathrm{FG}$ denotes the microscopic force acting on atom $i$ within the set of atoms $\mathcal{I}_I$ mapped to the CG site $I$.  

From the mapped atomistic trajectory, the effective CG interaction of OTP was then determined by utilizing Eq. (\ref{eq:forceresidual}) \cite{OpenMSCG}. In practice, the CG force field, as seen in Eq. (\ref{eq:forceresidual}), is expressed using pairwise basis sets $\phi_2 (R_{IJ})$ for the CG pair $I$ and $J$, resulting in
\begin{equation} \label{eq:pairwiseCG}
    \mathbf{F}_I (M_\mathbf{R}^N (\mathbf{r}^n);\phi_2)=\sum_{J\neq I} \phi_2(R_{IJ}) \hat{e}_{IJ},     
\end{equation}
where $\hat{e}_{IJ}$ denotes the unit vector of $\mathbf{R}_{IJ}$. As discussed in the main text, we utilize spline basis to enhance the expressivity of the basis sets. In practice, the B-spline function $\{ u_k\}$ with coefficients $c_k$ is introduced, i.e., $\phi_2(R_{IJ}) = \sum_k c_k u_k(R_{IJ})$. The effective CG force field can then be expressed as $\mathbf{F}_I (M_\mathbf{R}^N (\mathbf{r}^n);\phi_2)=\sum_{J \neq I}\sum_k c_k u_k (R_{IJ}) \hat{e}_{IJ}$, and Eq. (\ref{eq:forceresidual}) reduces to an overdetermined system of linear equations \cite{press2007numerical} of the following form:
\begin{equation} \label{eq:FM}
    \mathbf{F\phi}_2 = \mathbf{f}.
\end{equation}
Here, the matrix $\mathbf{F}$ represents the CG forces expressed through the overall CG configurations and basis functions $\mathbf{\phi}_2$, while $\mathbf{f}$ denotes the mapped atomistic forces at the CG level. Please refer to Ref. \onlinecite{lu2010efficient} for a more in-depth discussion of the interpolation scheme and implementation. 

\subsubsection*{2. Computational Details}
For the single-site representation of OTP molecules, force-matching was numerically performed using a spline resolution of 0.2 $\angstrom$ and by sampling the interaction up to a cutoff of 10.0 $\angstrom$ based on the reported characteristics of OTP interactions.

For the three-site (Lewis-Wahnstr\"{o}m model) representation, in order to account for flexible topology, we introduced fourth-order B-splines for the bonded interactions with a resolution of $0.20$ \angstrom. From manually mapped three-site trajectories of atomistic simulations, we observed that the bond length fluctuates from 4.09 to 4.65 \angstrom, supporting the initial hypothesis.

\subsection*{B. Bottom-up Connection to \textit{ad hoc} Lewis-Wahnstr\"{o}m Model}

\subsubsection*{1. Three-site CG Model}
From the bottom-up CG interaction depicted in Fig. \ref{fig:OTP-RDF},  understanding how coarse-graining a three-site Lennard-Jones representation to the single-site resolution renders a purely repulsive interaction, or vice versa, is not straightforward. This mismatch implies that the Lewis-Wahnstr\"{o}m model might not be the correct representation of microscopic OTP molecules at the three-site resolution. A systematic assessment of the fidelity of the Lewis-Wahnstr\"{o}m model in comparison with the microscopic reference is lacking in the literature. Therefore, we next performed coarse-graining of the atomistic OTP to the three-site resolution [Fig. \ref{fig:spatialCG}(b)] by mapping each benzene ring to the CG site. 

The original Lewis-Wahnstr\"{o}m model represents the OTP molecule as three particles interacting through a Lennard-Jones potential with $\varepsilon /k_B=600$ K (i.e., 1.1923 kcal/mol) and $\sigma=4.83$ \angstrom. The intramolecular interactions are constrained with a fixed bond length of $\sigma$ and a fixed angle of $75^{\circ}$. Although these fixed bonds and angles allow for exhibiting strong potential-virial correlations, this \textit{ad hoc} description does not align with chemical details and lacks microscopic details; the correct atomistic description of chemical bonds between benzene rings should be flexible. Additionally, in the Lewis-Wahnstr\"{o}m model, all three sites are considered equivalent with identical Lennard-Jones interactions, but the two CG particles at the end (\textit{tail}) should differ from the CG particle in the middle (\textit{center}), as the center CG site has one less hydrogen atom. This difference implies that tail-tail, center-center, and tail-center pairs should have distinct interaction profiles, at least in principle. 

To validate this \textit{ad hoc} description and gain deeper insight into the energetics at the single-site level, we propose a three-site CG model of OTP constructed from Fig. \ref{fig:OTP-interaction}(a) at $T=380$ K, $\rho=1.029$ g/ml. 
Similarly, the three-site spatial CG model was constructed by applying force-matching [Eq. (\ref{eq:forceresidual})] with bonded interactions. 
\begin{figure}
\includegraphics[width=1\columnwidth]{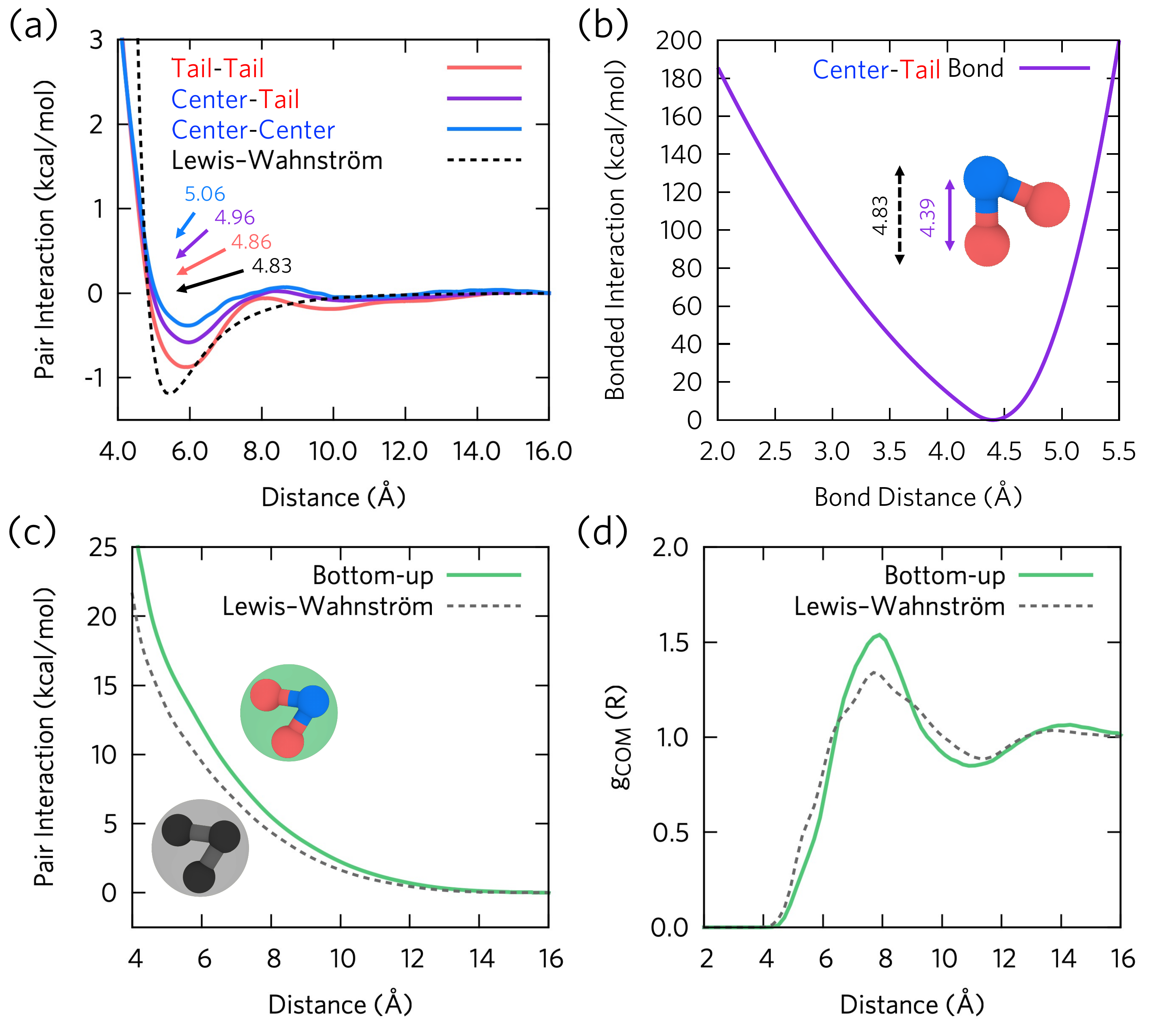}
\caption{\label{fig:OTP-LW} Assessing the \textit{ad hoc} Lewis-Wahnstr\"{o}m model (dotted lines) using the bottom-up three-site spatial CG model (solid lines). (a) Effective pair interactions. Bottom-up CG interactions exhibit a heterogeneous nature between tail and center CG sites with numbers indicating the zero interaction distance $R_0$, where $U(R_0)=0$. (b) Parametrized bond interactions from atomistic statistics, exhibiting a fluctuating nature, unlike the Lewis-Wahnstr\"{o}m model with 9.1\% shorter bond length. Despite these differences, (c) and (d) show that the purely repulsive interactions and pair correlations at the single-site level from the Lewis-Wahnstr\"{o}m model (black dots) are qualitatively consistent with the bottom-up one-site CG models (green lines), respectively.}
\end{figure}

\subsubsection*{2. Results}
For the non-bonded interactions, the tail and center CG sites depicted in Fig. \ref{fig:OTP-LW}(a) exhibit slightly different energetics but similar overall interaction profiles. Thus, we observe that the \textit{ad hoc} Lewis-Wahnstr\"{o}m model provides a good approximation of the site-site CG interactions. The qualitative agreement of the Lewis-Wahnstr\"{o}m model is attributed to its ability to capture short-range attractions. Yet, in a quantitative sense, the \textit{ad hoc} interaction is closest to the tail-tail interaction among the three pair interactions, with a difference of up to 0.2 kcal/mol. Due to the heterogeneity of CG sites, the center CG site exhibits less strong attraction, missing in the \textit{ad hoc} treatment. Additionally, the \textit{ad hoc} model is characterized by a single Lennard-Jones interaction with a long-range decay scaling as $R^{-6}$, whereas the bottom-up derived interactions show more complex decaying profiles with additional long-range valleys. 

In Fig. \ref{fig:OTP-LW}(b), in alignment with atomistic statistics, the parametrized CG bonded interaction exhibits a flexible, harmonic-like profile with a minimum of 4.39 \angstrom, which is shorter than the Lewis-Wahnstr\"{o}m value of $\sigma=4.83$ \angstrom. From this bond length, the mismatched bond interaction is approximately 30 kcal/mol ($\approx 40 k_B T$), indicating inaccuracies in the overall energetics during computer simulations. Taken together, Figs. \ref{fig:OTP-LW}(a) and (b) suggest that the Lewis-Wahnstr\"{o}m model can serve as a qualitative approximation for the accurate bottom-up CG model. The next step would be to examine whether the purely repulsive interaction profile at the single-site resolution is consistent with the description given by the Lewis-Wahnstr\"{o}m model.

To construct a single-site representation from the Lewis-Wahnstr\"{o}m model, we first generated the Lewis-Wahnstr\"{o}m system under the same temperature and density conditions (380 K and 1.029 g/ml). After placing 125 Lewis-Wahnstr\"{o}m molecules with correct bonding constraints, energy minimization using the stochastic descent algorithm was performed to eliminate artificial strains applied to the system. Then, we conducted the MD simulation while fixing the bonds and angles. Finally, from the propagated CG trajectories, three-site OTP models were mapped into a center-of-mass CG model, and then we used the same protocol described above to determine the effective Lewis-Wahnstr\"{o}m interaction at the single-site resolution. 

Figure \ref{fig:OTP-LW} compares the obtained single-site \textit{ad hoc} interaction with the spatial CG interaction. Notably, we still observe a purely repulsive interaction profile, even from the one constructed from the Lewis-Wahnstr\"{o}m model. In other words, for both analytic and phenomenological models, the coarse-graining of three Lennard-Jones sites with fixed topology can cancel out short-range attractions at the center-of-mass level. This leads to the important conclusion that the purely repulsive nature depicted in Fig. \ref{fig:OTP-interaction} is not merely a numerical artifact but is consistent with the phenomenological model. This conclusion is further substantiated by the center-of-mass RDF [Fig. \ref{fig:OTP-LW}(d)]. The single-site CG RDF of the Lewis-Wahnstr\"{o}m model shows a trend consistent with the bottom-up reference from atomistic simulations, where the slight differences can be attributed to the missing microscopic details in the \textit{ad hoc} treatment. 

To summarize, this agreement demonstrates the suitability of our model for rigorously studying the density scaling nature from a microscopic perspective and for achieving a more accurate representation of the OTP molecules by accounting for the heterogeneous nature of CG sites for non-bonded interactions, as well as for describing flexible bonded interactions.

\subsection*{C. Validation of Spatial Coarse-Graining}
\begin{figure} \label{fig:2pt-dos}
\includegraphics[width=0.75\columnwidth]{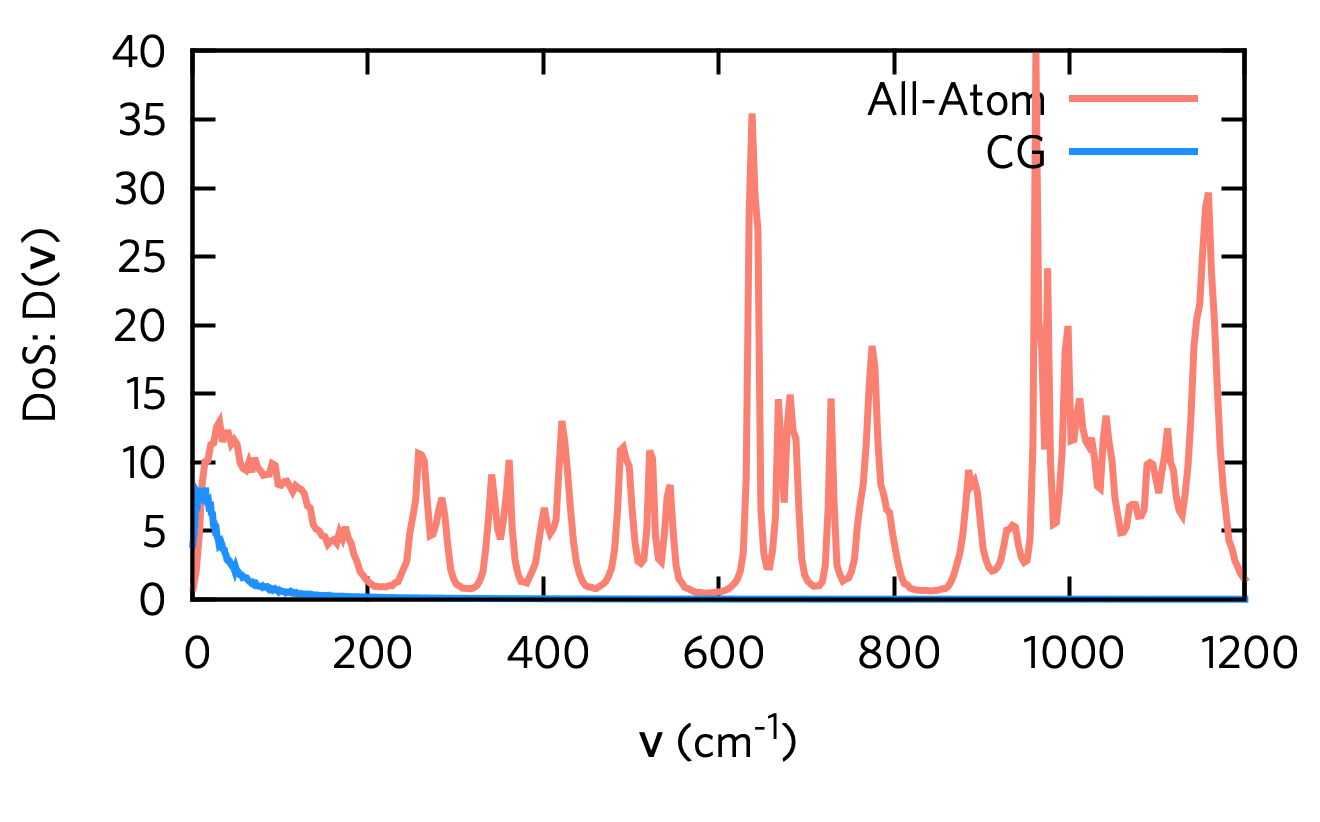}
\caption{\label{fig:OTP-PWR} Computed density of states from the velocity autocorrelation functions of atomistic (red) and CG (blue) OTP systems. The obtained DoS, $D(\nu)$, is plotted up to 1200 cm$^{-1},$ corresponding to the fingerprint region of a benzene ring from vibrational spectroscopy \cite{colthup2012introduction}.}
\end{figure}

We computed the power spectrum $D(\nu)$ from the velocity autocorrelation functions of the atomistic and CG models of OTP to demonstrate that the temporal CG approach can enhance the efficient exploration of the OTP molecules. The CG system only exhibits translational motions, i.e., $v_\mathrm{CG}=v_\mathrm{trans}$, whereas the atomistic molecules show both translation and vibrational motions: $v_\mathrm{FG} = v_\mathrm{trans}+v_\mathrm{rot}+v_\mathrm{vib}$. This difference is shown in Fig. \ref{fig:OTP-PWR}, where the CG model only exhibits translational motions at low frequencies below 100 cm$^{-1}$. Relatively high-frequency motions, which can be understood as the vibrational motions of OTP molecules, are all removed by coarse-graining. For example, we can identify the non-zero intensities of $D(\nu)$ near 500--1000 cm$^{-1}$ as the C-H vibrations of benzene \cite{colthup2012introduction}. A drastic quenching of these vibrational motions, mainly attributed to C-H vibrations, demonstrates the importance of spatial coarse-graining for exploring effective molecular motions in order to identify the virial-potential correlations of OTP. 

\subsection*{D. The Representability Issue in Spatial CG Models}
When estimating the scaling exponent for spatially CG models, the representability challenge introduced in Sec. \ref{subsec:representability} severely hampers the assessment of virial-potential correlations in CG models. Since the effective CG interaction is the free energy and not the internal energy \cite{noid2008multiscale,noid2008multiscale2}, the average CG energy estimated from CG simulations corresponds to the overall free energy at the fully atomistic (FG) level \cite{jin2019understanding}. Hence, potential energy fluctuations at the FG level cannot be directly assessed from the CG simulation. The situation is exacerbated when estimating the CG pressure because the na\"ively estimated CG virial and pressure differ significantly from their FG references \cite{dannenhoffer2019compatible}. This necessitates a reevaluation of the missing virial contribution throughout the CG simulation, making it highly impractical to evaluate correlations between the virial and potential energy simultaneously during the CG simulations. Therefore, unlike with temporal coarse-graining, this approach cannot be used to estimate $\gamma$. Furthermore, our bottom-up approach does not yield an analytical form of the CG potential; instead, it aims to find the closest discretized potential values at specific points defined by the spline resolution. Therefore, higher-order derivatives are not numerically stable and reliable, and $\gamma$ cannot be derived from the derivatives of analytical potentials, as with the IPL potential.

For the IPL pair potential following $V_\textrm{IPL}(R)=\epsilon (R/\sigma)^{-m}$, the density scaling exponent is constant over the radial domain as $\gamma=m/3$. Even for analytical potentials $V(R)$ that contain IPL-like hard-core repulsion, the effective distance-dependent exponent can be estimated from the $p$-th order derivatives of the analytical potential as $m^{(p)} = -p - R {V^{(p+1)}(R)}/{V^{(p)}(R)}$ \cite{scl_II}. However, this method is not applicable to bottom-up CG interactions parametrized according to Sec. \ref{subsec:CGSpace} because the CG potential is state point-dependent. Under spatial coarse-graining schemes, Eq. (\ref{eq:CGGamma}) instead provides a systematic approach to estimate the density scaling exponent.

\subsection*{E. Assessment of Lewis-Wahnstr{\"o}m Model: Thermodynamic Entropy}
In order to estimate the overall and intermolecular molar entropy of OTP, we selected the ambient temperature of 360 K under 1 atm conditions, corresponding to a box length of 35.80 \angstrom, as illustrated in Fig. 1(b). Employing the same atomistic simulation protocol, 2PT simulations were conducted to calculate the molar entropy.

First, from a fully atomistic simulation, the overall entropy was determined to be 350.54 J/mol/K. The translational and rotational (intermolecular) contributions at this state point amount to $S_\textrm{FG}^\textrm{trn}+S_\textrm{FG}^\textrm{rot} = 144.55$ J/K/mol. This indicates that more than half of the entropy arises from the vibrational contribution, underscoring the necessity for careful consideration of the missing entropy when determining the entropy of OTP. 

We evaluated the fidelity of the Lewis-Wahnstr{\"o}m model by constructing this approximate model for the same thermodynamic state point and conducted MD simulations for 2.5 ns. The 2PT estimation of entropy yielded 149.83 J/K/mol, almost identical to the intermolecular entropy of the atomistic OTP model. This agreement again substantiates our finding that the Lewis-Wahnstr{\"o}m model serves as a good approximation for the fully atomistic microscopic model of OTP. Furthermore, our estimated entropy of the Lewis-Wahnstr{\"o}m model is of the same order of magnitude as reported earlier in Ref. \onlinecite{mossa2002dynamics} using the inherent structure thermodynamic formalism \cite{stillinger1982hidden,stillinger1984packing} at $T = 360$ K and $T_0=5000$ K
\begin{align}
    S(T=360 K) & =S(T_0)+3R \log \left( \frac{T=360 K}{T_0} \right) \nonumber \\ &+ \int_{T_0}^{T=360 K} dT' \frac{1}{T'} \left(\frac{\partial U'(T')}{\partial T'}\right),
\end{align}
giving $S(360K)$ approximately 188.75 J/K/mol. 

\subsection*{F. Derivation of Eq. (\ref{eq:avramov_entropy})}
Equation (\ref{eq:avramov_entropy}) is derived from the total differential of $S$ (per particle) as a function of $T$ and $\nu$, $S(T,\nu)$, from $dS = \partial S/\partial T|_V dT +\partial S/\partial V|_T dV$:
\begin{equation}
    dS = \frac{c_v}{T}dT + \left( \frac{c_p-c_v}{V\alpha_p T}\right)dV,
\end{equation}
where we use the thermodynamic relationship between $c_p$ and $c_v$:
\begin{equation} \label{eq:pp}
    c_p = c_v + TV \alpha_p \left(\frac{\partial P}{\partial T} \right)_v.
\end{equation}
Assuming that $c_v$ and $c_p-c_v$ are almost constant over varying temperatures and volumes, we can integrate $S(T,\nu)$ from a reference state $S_\mathrm{ref}$, yielding 
\begin{equation} \label{eq:avramov}
    S(T,\nu) = S_\mathrm{ref}+c_v \left[\ln \left(\frac{T}{T_\mathrm{ref}} \right) + \frac{c_p-c_v}{\alpha_P Tc_v} \ln \left(\frac{\nu}{\nu_\mathrm{ref}} \right)\right].
\end{equation}
This approximation holds reasonably well over modest ranges of temperature and pressure \cite{casalini2006thermodynamic}.

As discussed in the main text, we chose the reference state as an ideal gas-like state, differing from the conventional Avramov model that uses the glass transition point. A detailed discussion on determining the reference state is provided in Appendix G. Now, defining $\gamma_G = (c_p/c_v-1)/(\alpha_p T)$, which approximates $\gamma_G\approx \gamma$  \cite{casalini2006thermodynamic}, we arrive at Eq. (\ref{eq:avramov_entropy}).

\subsection*{G. Estimation of Additional Thermodynamic Properties}

\subsubsection*{1. Estimation of the Progogine-Defay Ratio}
In order to compute the thermodynamic Prigogine-Defay ratio, defined as \cite{chemical_thermodynamics, Gupta1976, Takahara1999, Schmelzer2006, Ellegaard2007, Pedersen2008b, Gundermann2011, Fragiadakis2011, Casalini2011, Tropin2012, Garden2012}
\begin{equation} \label{eq:Prigo}
    \Pi\equiv\frac{\Delta c_p\Delta \kappa_T}{V_gT_g(\Delta \alpha_p)^2},
\end{equation}
we conducted constant $NpT$ simulations at 1 atm for a 200 ns simulation (following 67 ns of equilibration) at different temperatures as illustrated in Fig. \ref{fig:prigogine_defay}. The glass transition, identified by changes in slope in Fig. \ref{fig:prigogine_defay}(a), was observed on the simulation time scale. Systems near the glass transition that showed partial relaxation (unfilled circles in Fig. \ref{fig:prigogine_defay}) were excluded from further analysis. The specific volume at $T_g$ was determined as $V_g=219(2)$ cm$^3$/mol.

At the glass transition temperature, $\Pi$ was computed from jumps in response functions, utilizing the following equations \cite{Allen1987}. The variance of instantaneous enthalpy ($\mathcal{H}+pV$) fluctuations, i.e., $\sigma^2(\mathcal{H}+pV):=\langle \Delta(\mathcal{H}+pV)^2\rangle$,
\begin{equation} \label{eq:prigocp}
    \langle (\Delta (\mathcal{H}+pV))^2 \rangle_{NpT} = k_B T^2 c_p,
\end{equation}
was used to estimate the isobaric heat capacity $c_p$. Then, the variance of the volume fluctuations
\begin{equation}
    \langle (\Delta V)^2 \rangle_{NpT} = Vk_B T \kappa_T,
\end{equation}
was used to estimate the isothermal compressibility $\kappa_T$, and the covariance between volume and enthalpy
\begin{equation}
    \langle \Delta V \Delta (\mathcal{H}+pV)\rangle_{NpT} = k_BT^2 V\alpha_p,
\end{equation}
was used to obtain the thermal expansion coefficient $\alpha_p$. From the microscopic OTP simulations, the steps in these response functions at $T_g$ were found as $\Delta c_p=86(5)$ J/mol$\cdot$K, $\Delta \kappa_T=1.75(10)\times10^{-10}$ Pa$^{-1}$, and $\Delta \alpha_p=4.42(10)\times10^{-4}$\ K$^{-1}$, yielding $\Pi=1.02(10)$ near unity using Eq. (\ref{eq:Prigo}). Figure \ref{fig:prigogine_defay} further illustrates these jumps in the response functions with the error (i.e., one standard deviation) estimated from Monte Carlo sampling, assuming Gaussian noise.

\subsubsection*{2. Scaling Slope Estimation}
\begin{figure} 
\includegraphics[width=0.55\columnwidth]{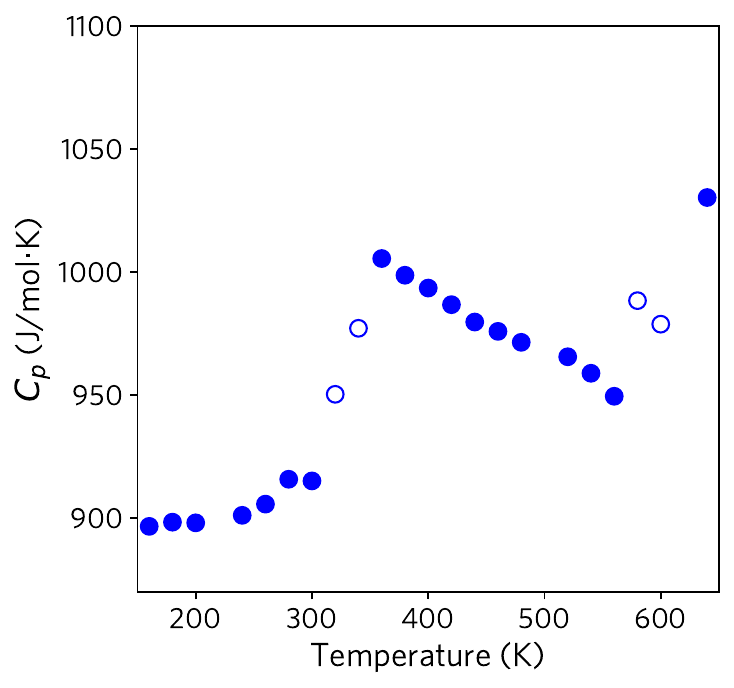}
\caption{\label{fig:OTP-CP} Determination of the reference state at $600$ K for estimating $S_\mathrm{ref}$ based on the specific isobaric heat capacity, $c_p$ (J/mol$\cdot$K).}
\end{figure}

We first estimated the reference state near the boiling point based on the jumps of the specific isobaric heat capacity, using the same computational protocol as in Eq. (\ref{eq:prigocp}), see Fig. \ref{fig:OTP-CP}.
Next, we estimated $c_v$ from the computed $c_p$, $\alpha_p$, and $\kappa_T$ values based on Eq. (\ref{eq:pp})
\begin{equation}
    c_v = c_p - \frac{TV\alpha_p^2}{\kappa_T}.
\end{equation}

\bibliography{manuscript_v2}

\end{document}


\title{Supplemental Material: Microscopic Theory of Density Scaling: Coarse-Graining in Space and Time}
\author{Jaehyeok Jin}
\email{jj3296@columbia.edu}
\affiliation{Department of Chemistry, Columbia University, New York, New York 10027, United States}
\author{David R. Reichman}
\email{drr2103@columbia.edu}
\affiliation{Department of Chemistry, Columbia University, New York, New York 10027, United States}
\author{Jeppe C. Dyre}
\email{dyre@ruc.dk}
\affiliation{``Glass and Time'', IMFUFA, Department of Science and Environment, Roskilde University, P.O. Box 260, DK-4000 Roskilde, Denmark}
\author{Ulf R. Pedersen}
\email{ulf@urp.dk}
\affiliation{``Glass and Time'', IMFUFA, Department of Science and Environment, Roskilde University, P.O. Box 260, DK-4000 Roskilde, Denmark}\date{\today}

\maketitle
\section{Calculation Details}
\subsection{Atomistic OTP Simulations}
We used the Optimized Potentials for Liquid Simulations (OPLS) force field with charge corrections, 1.14*CM1A(-LBCC) \cite{Dodda2017b}. In detail, the overall potential energy of the OTP force field is represented as:
\begin{equation}\label{eq:energy_surface}
    U({\bf r}^n) = U_\mathrm{intra}({\bf r}^n) + U_\mathrm{inter}({\bf r}^n),
\end{equation}
where the molecular shape is retained by intramolecular bonds, angles, and dihedral potential, $U_\textrm{intra}({\bf r}^n)$, while the intermolecular interactions, $U_\textrm{intra}({\bf r}^n)$, are described by Coulomb and Lennard-Jones interactions. In particular, $U_\textrm{intra}({\bf r}^n)$ can be represented as:
\begin{align}\label{eq:energy_surface_intra}
    U_\mathrm{intra}({\bf r}^n) & = \sum_\mathrm{bonds}K_r(r_{ij}-r_\mathrm{eq})^2+\sum_\mathrm{angles}K_\theta(\theta_{ijk}-\theta_\mathrm{eq})^2 \nonumber \\ &+ \sum_\mathrm{dihedrals}V(\phi_{ijkl})
\end{align}
with $V(\phi_{ijkl})=\sum_{n=1}^3 V_n\cos(n\phi_{ijkl}+f_n)/2$, and 
\begin{equation}\label{eq:energy_surface_inter}
U_\mathrm{inter}({\bf r}^n) = \sum_{i>j}\frac{q_iq_je^2}{r_{ij}}+\frac{A_{ij}}{r_{ij}^{12}}-\frac{C_{ij}}{r_{ij}^6}
\end{equation}
with $r_{ij}=|{\bf r}_j-{\bf r}_i|$ as a pair distance for the intermolecular interactions. The parameters ($K_r$, $r_\mathrm{eq}$, $K_\theta$, $\theta_\mathrm{eq}$, $V_n$, $f_n$, $q_i$, $A_{ij}$ and $C_{ij}$) are chosen according to Ref.\ \onlinecite{Dodda2017b}.

\subsection{Self-Diffusion Coefficients}
In order to perform density scaling from atomistic OTP simulations in Fig. 15, we first computed the self-diffusion coefficient from the long-time limit of the atomistic mean square displacement using Einstein's relation 
\begin{equation}
    D = \lim_{t\to\infty} \langle (\Delta r (t))^2\rangle/6t,
\end{equation}
where $\Delta r(t)=|{\bf r}(t)-{\bf r}(0)|$ is the magnitude of the displacement of a given atom within a time-interval $t$, and the average $\langle\ldots\rangle$ is taken over all the atoms and initial times.

\section{Temporal CG Model of OTP: Correlation}
\begin{figure*}
\includegraphics[width=0.35\columnwidth]{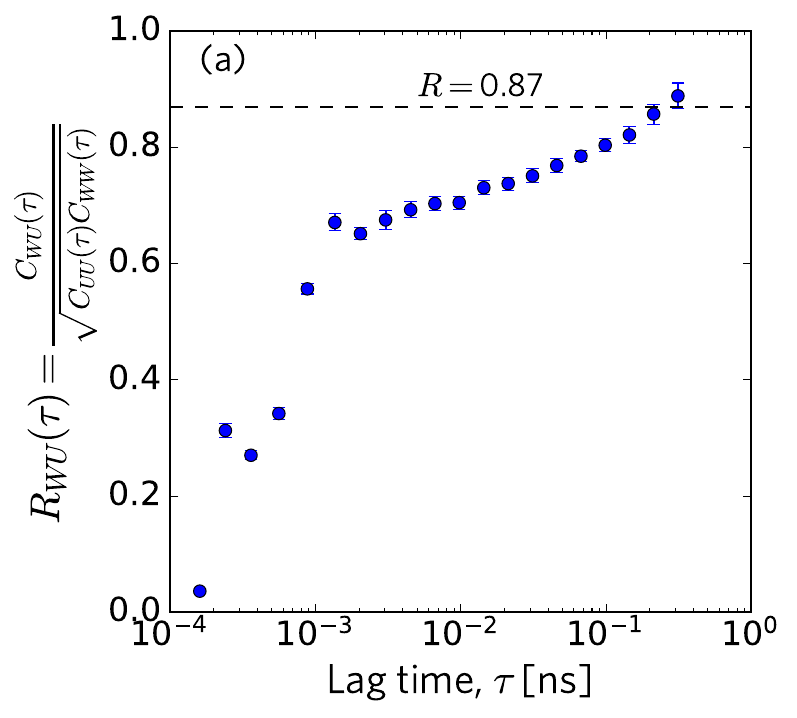}
\includegraphics[width=0.33\columnwidth]{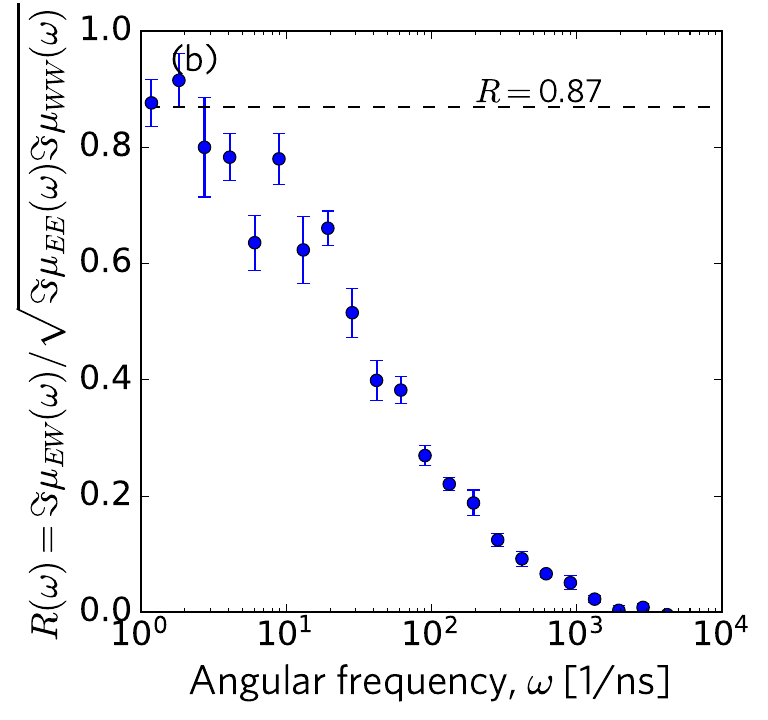}
\caption{\label{fig:rvalue} The Pearson correlation coefficient of the time-averaged potential energy and virial using (a) the time correlation function approach $R_{WU}(\tau)$ and (b) the frequency-dependent response approach $R(\omega)$.}
\end{figure*}

Similar to Fig. 6(a) in the main text, which illustrates the time-averaging approach, this section provides supplementary results for other temporal coarse-graining methods. Figure \ref{fig:rvalue} delineates the computed Pearson coefficients for the time correlation function approach [Fig. \ref{fig:rvalue}(a)] and frequency-dependent response approach [Fig. \ref{fig:rvalue}(b)].

\section{Spatial CG Model of OTP: Single-Site Resolution}
\subsection{CG Pair Interactions}
\begin{figure*}
\includegraphics[width=0.75\columnwidth]{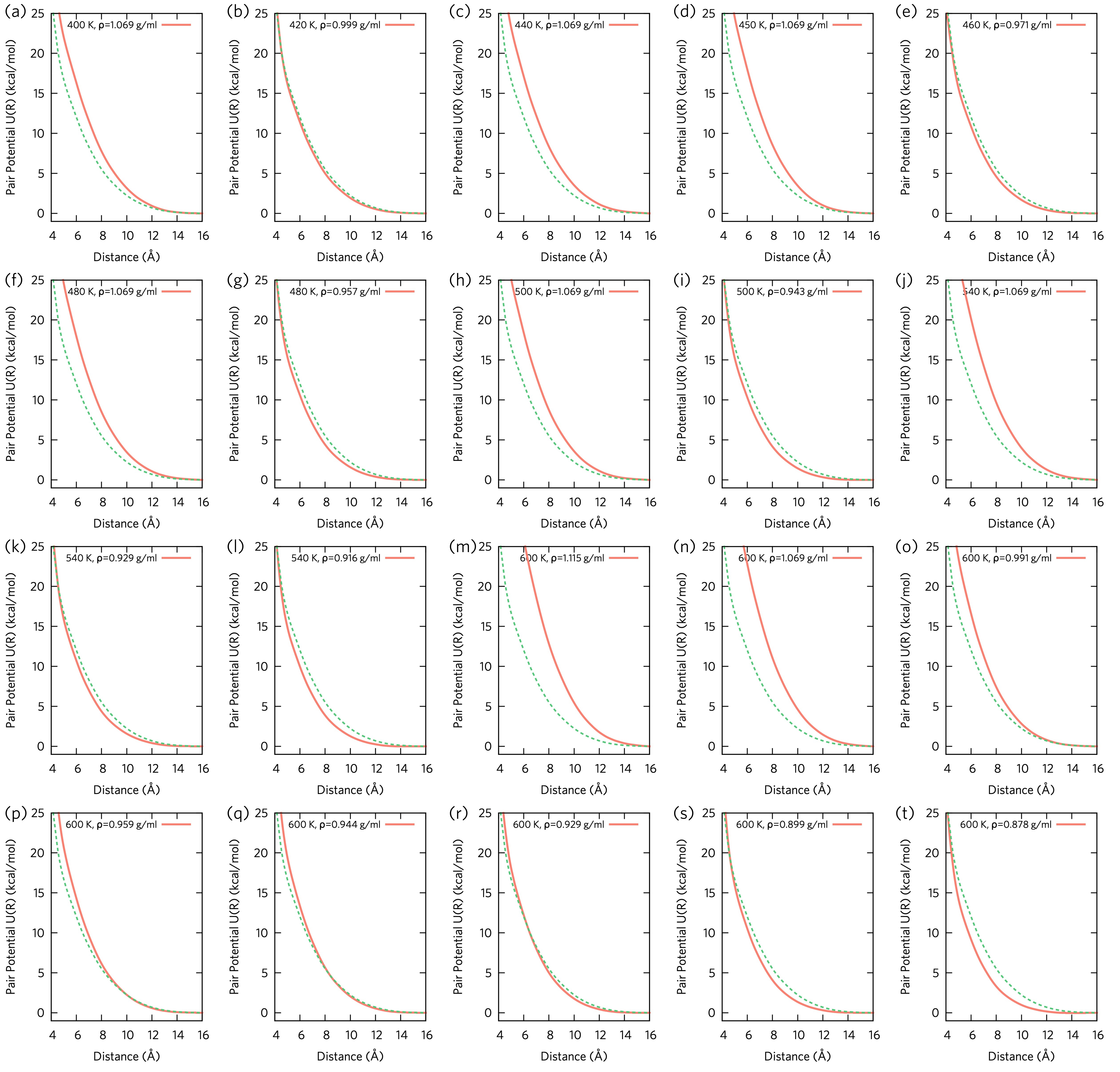}
\caption{\label{fig:OTP-interaction2} Parametrized pair interactions of the single-site CG OTP at various temperatures and densities using force-matching for the state points except for the one depicted in Fig. 10. Temperature and density conditions are denoted in the title of each panel. For clarity, the CG interaction from Fig. 10(a) is plotted as a reference (green dots) against each CG interaction (red lines).}
\end{figure*}

To further illustrate the complex nature of bottom-up CG interactions and emphasize the challenges in their inference, Fig. \ref{fig:OTP-interaction2} depicts how pair interactions for the spatial CG model of OTP vary with density and temperature. The general trend aligns with our discussion pertinent to Fig. 10, but notably, there is no underlying unified potential in a reduced unit, as is the case with simple analytical interactions. This is because the bottom-up CG interaction includes an entropic contribution, whereas conventional interactions for point particles are purely energetic. For instance, at a fixed volume (constant $NVT$), changes in pair interaction can be understood as the entropic contribution to the CG PMF, as demonstrated by studies of CG liquids. However, the exact forms of entropic and energetic terms for molecular systems involve high-dimensional integrals of FG degrees of freedom, making them challenging to determine.

\subsection{CG Pair Correlations}
\begin{figure*}
\includegraphics[width=0.75\columnwidth]{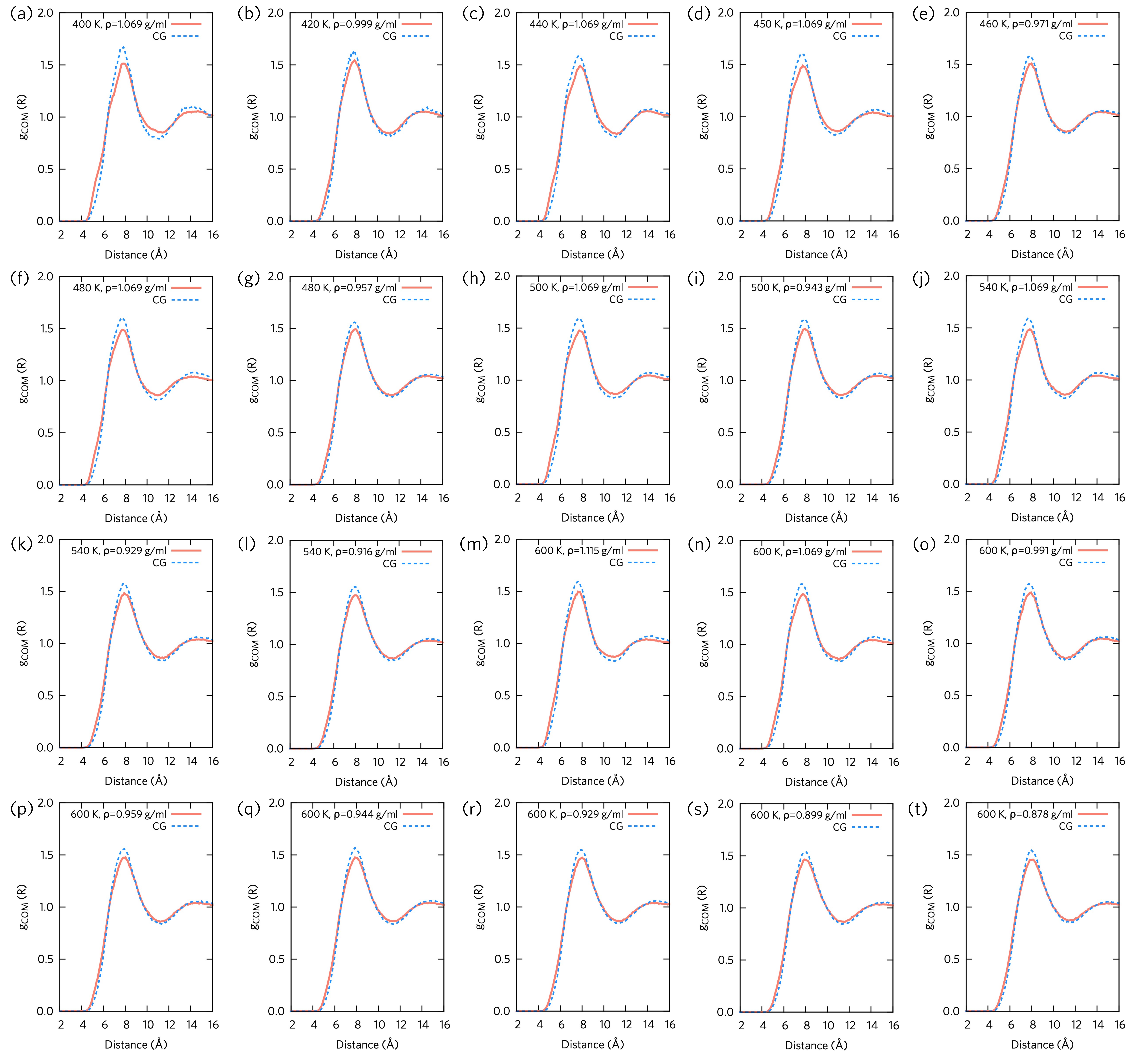}
\caption{\label{fig:OTP-rdf2} Center-of-mass pair correlation functions $g_\mathrm{COM}(R)$ for atomistic (red lines) and the single-site CG (blue dots) simulations of OTP at various temperatures and densities using force-matching for the state points except for the one depicted in Fig. 11. Temperature and density conditions are denoted in the title of each panel.}
\end{figure*}

Similar to Fig. 11 in the main text, Fig. \ref{fig:OTP-rdf2} provides a comprehensive comparison of the pair correlation function between the spatial CG models and the atomistic reference. As discussed in the main text, interestingly, RDFs over different conditions do not seem to differ significantly with drastically different dynamics (as plotted in Fig. 15). This confirms our assertion that higher-order contributions beyond pair excess entropy should be taken into account when computing excess entropy. Our 2PT-based approach was able to distinguish different excess entropy values across these conditions.

\section{Excess Entropy Scaling: Atomistic OTP}
\subsection{Computational Details}
In order to compute the complete excess entropy of the fully atomistic OTP, we need to calculate rotational and vibrational excess entropies in addition to the translational excess entropy that was already explained in the main text. As discussed in Appendix B, for non-linear molecules, this necessitates determining 3 rotational temperatures and $3N-6=32\times3-6=90$ vibrational temperatures to utilize Eqs. (32) and (33) from the main text. While this can be done practically for relatively simple liquids by referring to the experimental spectrum, obtaining the full rotational and vibrational spectrum for OTP is relatively challenging  \cite{baranovic2001molecular}. Therefore, in this work, we approximate these rotational and vibrational temperatures using the quantum mechanical \textit{ab initio} calculations conducted at the B3LYP/6-31G* level of theory. Complete frequencies are publicly available through the NIST Chemistry WebBook for the OTP molecule \cite{otp-nist}.

\subsection{Rotational Excess Entropy}
In Ref. \onlinecite{otp-nist}, the rotational constants of OTP were obtained as shown in Table \ref{table:rot}, where $\nu_A>\nu_B>\nu_C$ without loss of generality. In order to employ Eq. (32), we need to convert $\nu_A$, $\nu_B$, and $\nu_C$ to the characteristic rotational temperature, i.e., $\Theta_{A,B,C}$. This can be readily done by
\begin{equation}
    \Theta_{A} = \frac{h \nu_A}{k_B} = 0.0238 \,\mathrm{(K)}
\end{equation}
for $A=0.49701$ (GHz). 

Additionally, we note that for OTP the rotational symmetry number $\sigma$ is 2 based on the point group symmetry and Ref. \onlinecite{shimanouchi1978tables}. Therefore, by combining $\Theta_A$, $\Theta_B$, and $\Theta_C$ with Eq. (32), we arrive at
\begin{equation}
    s_{id}^\mathrm{rot} = 6.6930 + \frac{3}{2} \ln T.
\end{equation}

\begin{table}
\caption{Rotational constants and rotational temperatures of OTP \cite{otp-nist}} \label{table:rot}
\begin{tabular}{lrr}
\toprule
 & \textbf{Value $\nu$ (GHz)} & \textbf{Temperature $\Theta$ (K)}\\
\midrule
A & 0.49701 & 0.02385\\
B &  0.42977 & 0.02063\\
C & 0.25663 & 0.01232\\
\bottomrule
\end{tabular}
\end{table}

\subsection{Vibrational Excess Entropy}
Similarly, the IR frequencies $\tilde{\nu}$ (in cm$^{-1}$) reported in Ref. \onlinecite{otp-nist} are summarized in Table  \ref{table:vib}. Since the total vibrational degrees of freedom are 90, we envision that the vibrational contribution to the excess entropy should not be disregarded, unlike simple liquids, e.g., water studied in Ref. \onlinecite{jin2023understanding}. 

\begin{table}[h]
    \caption{IR frequencies $\tilde{\nu}$ (cm$^{-1}$) and intensities of OTP \cite{otp-nist}}
    \label{table:vib}
    \centering
    \begin{tabularx}{\textwidth}{*{6}{>{\centering\arraybackslash}X}}
        \toprule
        \textbf{Frequency (cm$^{-1}$)} & \textbf{Intensity (km/mol)} &
        \textbf{Frequency (cm$^{-1}$)} & \textbf{Intensity (km/mol)} &
        \textbf{Frequency (cm$^{-1}$)} & \textbf{Intensity (km/mol)} \\
        \midrule
        52.6642  & 0.0215 & 135.5863 & 0.1414 & 515.1485 & 2.1844 \\
        53.7055  & 0.0714 & 235.5080 & 0.0958 & 532.8955 & 3.5319 \\
        63.1924  & 0.0378 & 256.5623 & 0.0472 & 569.8793 & 7.4159 \\
        97.0696  & 0.0059 & 275.6565 & 0.0009 & 571.5684 & 0.1017 \\
        104.3725 & 0.8006 & 323.8709 & 0.3124 & 625.7981 & 5.1025 \\
        365.1529 & 0.0288 & 401.5943 & 0.4472 & 629.5127 & 1.8970 \\
        418.4642 & 0.0510 & 419.2888 & 0.3347 & 636.1307 & 0.2929 \\
        638.6079 & 2.3172 & 714.7017 & 8.7170 & 755.4640 & 4.4328 \\
        716.6825 & 44.0175 & 724.0355 & 1.0195 & 765.6536 & 70.3977 \\
        787.2457 & 1.7076 & 791.1876 & 9.4970 & 793.0835 & 9.9940 \\
        861.5338 & 0.1498 & 862.5151 & 0.2627 & 893.6913 & 0.0056 \\
        929.5802 & 4.4765 & 933.3170 & 1.4605 & 958.2162 & 1.1443 \\
        969.3511 & 0.1365 & 970.2425 & 0.3759 & 987.2366 & 0.1216 \\
        994.8300 & 0.5755 & 995.3742 & 0.2520 & 1017.1845 & 1.5351 \\
        1024.2037 & 5.3050 & 1026.6683 & 4.2950 & 1064.8196 & 0.8206 \\
        1065.7236 & 0.4210 & 1094.0429 & 1.7227 & 1109.1670 & 6.2549 \\
        1111.8772 & 0.0869 & 1146.3365 & 0.4583 & 1192.5458 & 0.1006 \\
        1192.9468 & 0.0010 & 1197.5335 & 0.1546 & 1215.5277 & 1.4452 \\
        1216.3675 & 0.0082 & 1274.8493 & 0.3482 & 1292.5816 & 0.9630 \\
        1312.1516 & 3.0332 & 1325.2794 & 0.0282 & 1340.8326 & 0.5582 \\
        1342.2142 & 0.1165 & 1365.5181 & 0.7908 & 1367.7039 & 0.0239 \\
        1474.2767 & 8.8966 & 1488.2231 & 4.9074 & 1499.8277 & 4.1337 \\
        1523.7133 & 26.3692 & 1545.0024 & 7.2483 & 1551.4151 & 0.9831 \\
        1617.5843 & 0.0199 & 1633.5802 & 2.8723 & 1638.5179 & 0.2614 \\
        1654.5157 & 0.9056 & 1661.3932 & 4.2452 & 1661.5755 & 4.1462 \\
        3178.9076 & 3.0931 & 3179.3764 & 3.0274 & 3182.6472 & 0.3958 \\
        3186.7503 & 3.1255 & 3187.1500 & 5.2363 & 3191.9456 & 12.8844 \\
        3196.2293 & 23.1248 & 3196.7760 & 21.0754 & 3201.4528 & 17.4704 \\
        3205.6309 & 53.3075 & 3205.9196 & 26.6773 & 3210.8177 & 31.7009 \\
        3213.8722 & 7.7278 & 3214.3598 & 8.3177 & & \\
        \bottomrule
    \end{tabularx}
\end{table}

From the reported IR frequency, the characteristic temperature for each $j$-th vibrational mode $v_j$ can be estimated as 
\begin{equation}
    v_j = \frac{ h \tilde{\nu} c}{k_B}.
\end{equation}
This conversion yields a temperature of $1.44$ K equivalent to the frequency of 1 cm$^{-1}$. For reference, the analysis code for converting all vibrational frequencies to the vibrational temperatures and computing the ideal gas entropy part based on 
\begin{equation}
    s_{id}^\mathrm{vib} = \sum_{j=1}^{90} \left[ \frac{\Theta_{v_j}/T}{e^{\Theta_{v_j}/T}-1} - \ln \left(1-e^{-\Theta_{v_j}/T} \right)\right]
\end{equation}
is available in the public repository \cite{zenodo}. In summary, a numerical estimation provides $s_{id}^\mathrm{vib}$ in Table \ref{table:vib2} at different temperatures.

\begin{table}[h]
    \caption{Ideal gas vibrational entropy (dimensionless) of OTP at different temperatures studied in this work}
    \label{table:vib2}
    \centering
    \begin{tabular}{cccc}
    \toprule
        \textbf{Temp (K)} & \(s_{id}^\mathrm{vib}\) & \textbf{Temp (K)} & \(s_{id}^\mathrm{vib}\) \\
        \midrule
        380 & 28.1916 & 460 & 35.2265 \\
        400 & 29.9455 & 480 & 36.9825 \\
        420 & 31.7050 & 500 & 38.7317 \\
        440 & 33.4664 & 540 & 42.2020 \\
        450 & 34.3468 & 600 & 47.3134 \\
        700 & 55.5185 & & \\
        \bottomrule
    \end{tabular}
\end{table}

\subsection{Excess Entropy Scaling for OTP}
Combined together, the excess entropy of OTP systems was evaluated by numerically estimating
\begin{align}
        s_\textrm{ex}^\mathrm{FG} & = s_\textrm{ex}^\mathrm{trn}+s_\textrm{ex}^\mathrm{rot}+s_\textrm{ex}^\mathrm{vib} \nonumber \\ 
                       & = s^\mathrm{FG} - \left[\frac{5}{2} - \ln \left( \frac{h^2 }{2\pi m k_B T}\right)^\frac{3}{2} -\ln \left( \frac{N}{V} \right)\right]  -\ln \left[ \frac{\sqrt{\pi}}{\sigma}\left( \frac{T^3 e^3}{\Theta_A \Theta_B \Theta_C}\right)^\frac{1}{2}\right]  - \sum_{j=1}^{3N-6} \left[ \frac{\Theta_{v_j}/T}{e^{\Theta_{v_j}/T}-1} - \ln \left(1-e^{-\Theta_{v_j}/T}\right)\right].
\end{align}
The evaluated values for each state point are available in the uploaded repository \cite{zenodo}. 

\section{Coarse-graining the Lennard-Jones model in time}
\subsection{Settings}
To demonstrate the fidelity of the proposed CG methodology beyond the OTP molecule, which only exhibits purely repulsive interaction at the CG level, we apply the temporal coarse-graining approach to a single-component Lennard-Jones system. Since the Lennard-Jones interaction contains both hard-core repulsion and long-range attraction described by the potential function
\begin{equation}
    V_{LJ}(R)=4\varepsilon\left[ \left(\frac{\sigma}{R}\right)^6-\left(\frac{\sigma}{R}\right)^{12}\right],
\end{equation}
this extension serves as a prototypical demonstration of complex molecules. In practice, the initial setup for the Lennard-Jones simulation was prepared by placing $N=4000$ particles at a number density of $\rho=0.844\sigma^{-3}$ and a temperature of $T=0.667k_B/\varepsilon$, corresponding to 35.7 mol/L and 80 K in Argon units with $\sigma=0.34$ nm, $\varepsilon=0.997$ kJ/mol, and $m=39.948$ u. Note that we also truncated and shifted $V_{LJ}(R)$ at 2.5$\sigma$. Lennard-Jones simulations were conducted using the LAMMPS software package, utilizing the same version as the OTP case. We sampled the MD configuration under the canonical ensemble using a Langevin thermostat (Grønbech-Jensen/Farago \cite{gronbech2013simple, gronbech2014}) with a time step of 0.005$\sigma\sqrt{m/\varepsilon}$ (11 fs in Argon units) and a thermostat coupling time of $\sigma\sqrt{m/\varepsilon}$ (2.2 ps in Argon units). After equilibration, the MD trajectory was propagated for $10^6$ steps, corresponding to 10.78 ns in Argon units. 

\subsection{Results}
\begin{figure*}
\includegraphics[width=0.45\columnwidth]{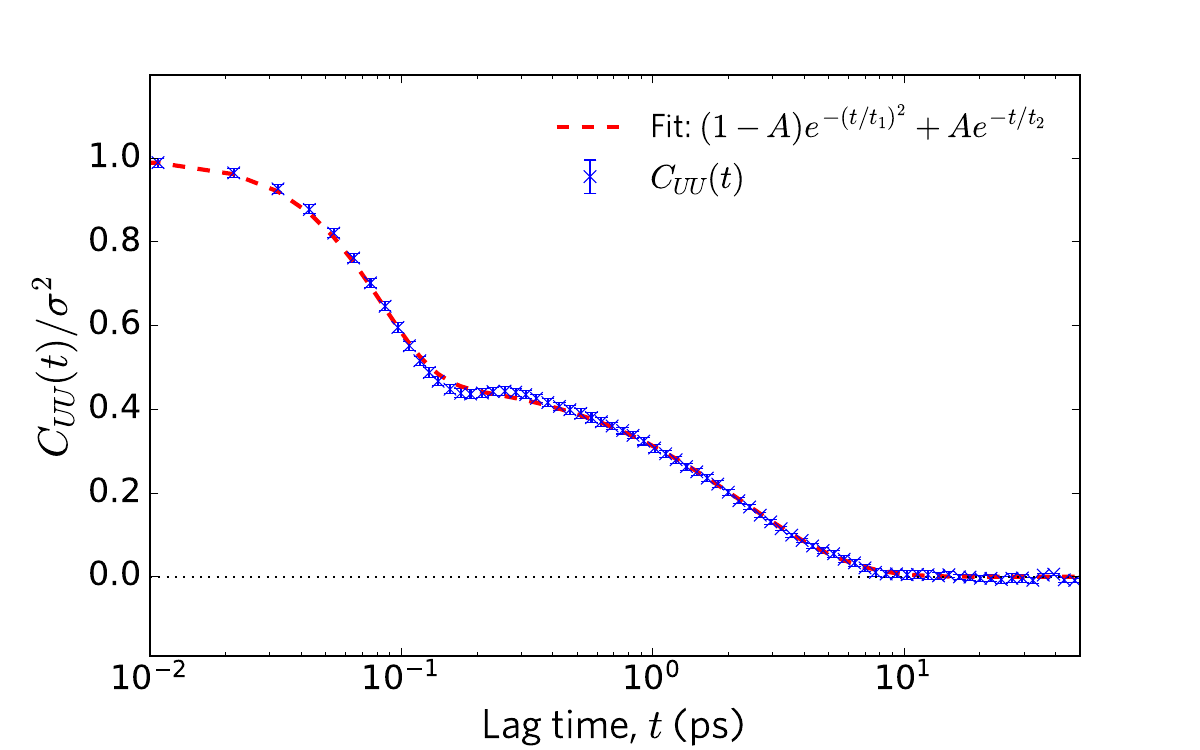}
\caption{\label{fig:lennard_jones_CUU} Potential energy auto-correlation function, $C_{UU}(t)$, normalized by the variance $\sigma^2$. The energy fluctuation shows relaxation on two timescales. The correlation function is fitted to $(1-A)\exp(-(t/t_1)^2)+A\exp(-(t/t_2))$, yielding $A=0.483(13)$, $t_1=83.1(10)$ fs, and $t_2=2.32(7)$ ps, by minimizing the sum of the squared residuals. The first relaxation ($t_1$) is related to neighbor collisions, and the second relaxation ($t_2$) is related to the Langevin thermostat.}
\end{figure*}

We now present our results for coarse-graining in time in Argon units. To provide a quantitative measure, we estimated the error bar by dividing the simulation trajectory into statistically independent blocks, following the approach outlined in the main text \cite{Flyvbjerg1989}. First, we analyzed the auto-correlation function of potential energy fluctuations, $C_{UU}(t)$, as depicted in Fig. \ref{fig:lennard_jones_CUU}. By fitting $C_{UU}(t)$ to $(1-A)\exp(-(t/t_1)^2)+A\exp(-t/t_2)$, we observed that the potential energy decorrelates on two characteristic timescales: (i) a fast timescale related to particle collisions, $t_1=83$ fs, and (ii) a slower timescale, $t_2=2.3$ ps, related to the Langevin thermostat.

\begin{figure*}
\includegraphics[width=0.35\columnwidth]{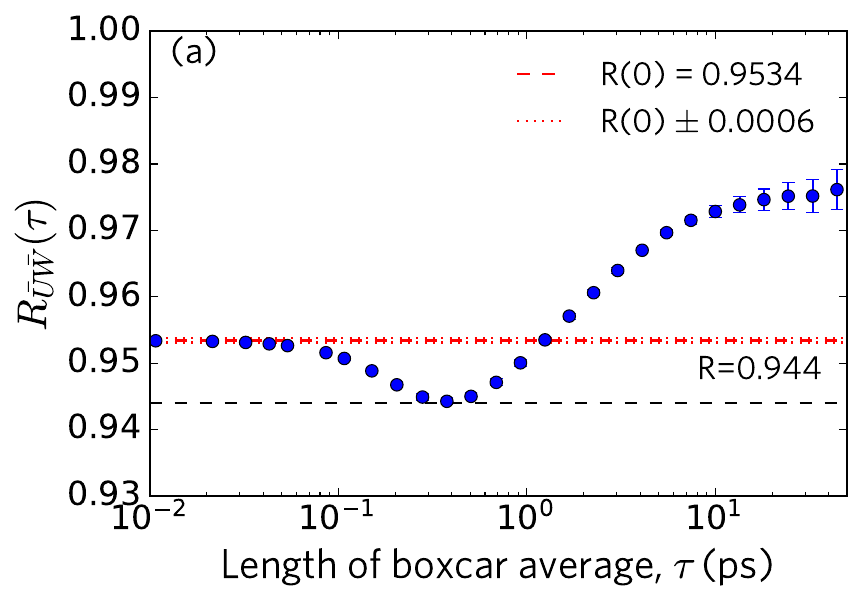}
\includegraphics[width=0.35\columnwidth]{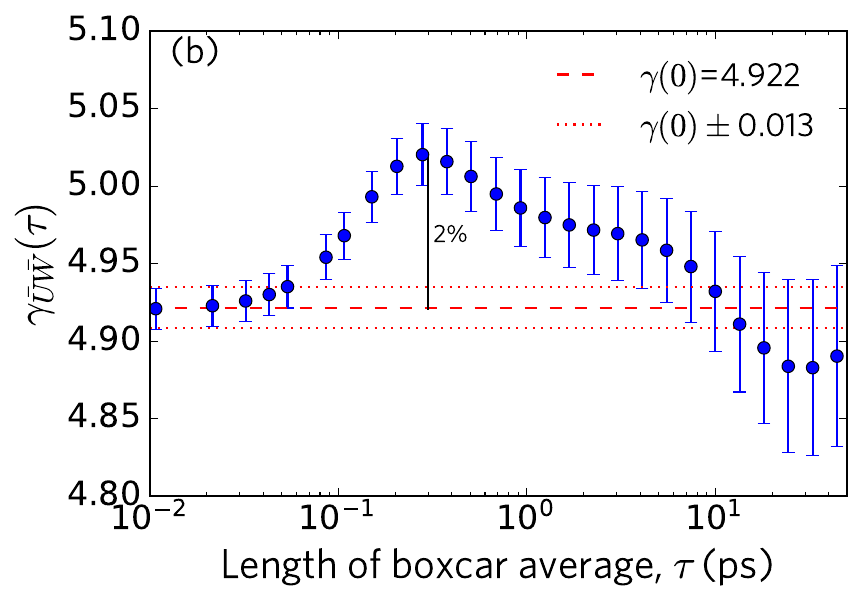}
\caption{\label{fig:lennard_jones_corr} Correlation coefficients from time-averaged CG Lennard-Jones. (a) The Pearson correlation coefficient $R_{\bar W\bar U}(\tau)$ of the boxcar averaged virial ($\bar W$) and potential energy ($\bar U$). (b) Density scaling exponent $\gamma_{\bar U\bar W}(\tau)$ for a range of time, $\tau$, (dots). 
}
\end{figure*}

From the auto-correlation analysis of the Lennard-Jones system, we note that the OTP molecules exhibit nearly uncorrelated instantaneous $WU$ fluctuations, whereas, for the Lennard-Jones system, we found strong correlations when performing time-averaging over the trajectory by averaging the energy and virial from $t$ to $t+\tau$ with equal weights, i.e., boxcar average. This phenomenon can be attributed to a hidden scale invariance of the slowly relaxing coordinates of the molecular system. 

Moving forward, we evaluate the scale invariance of the temporal CG Lennard-Jones system. Similar to the OTP case, we first assessed the Pearson correlation coefficient, $R_{\bar U \bar W}(\tau)$, from the temporally averaged virial ($\bar W$) and potential energy ($\bar U$) fluctuations. Figure \ref{fig:lennard_jones_corr}(a) confirms our hypothesis by demonstrating strong correlations in the slow energy-virial fluctuations resulting from a slow Lennard-Jones relaxation. Specifically, we find that $R(t)$ is above 0.944 for all characteristic times $\tau$. We then computed the density scaling coefficient from temporal coarse-graining, defined as 

\begin{equation}
    \gamma_{\bar U \bar W} = \frac{\langle \Delta \bar W \bar U\rangle}{\langle (\Delta \bar U)^2\rangle}.
\end{equation}

\begin{figure*}
\includegraphics[width=0.45\columnwidth]{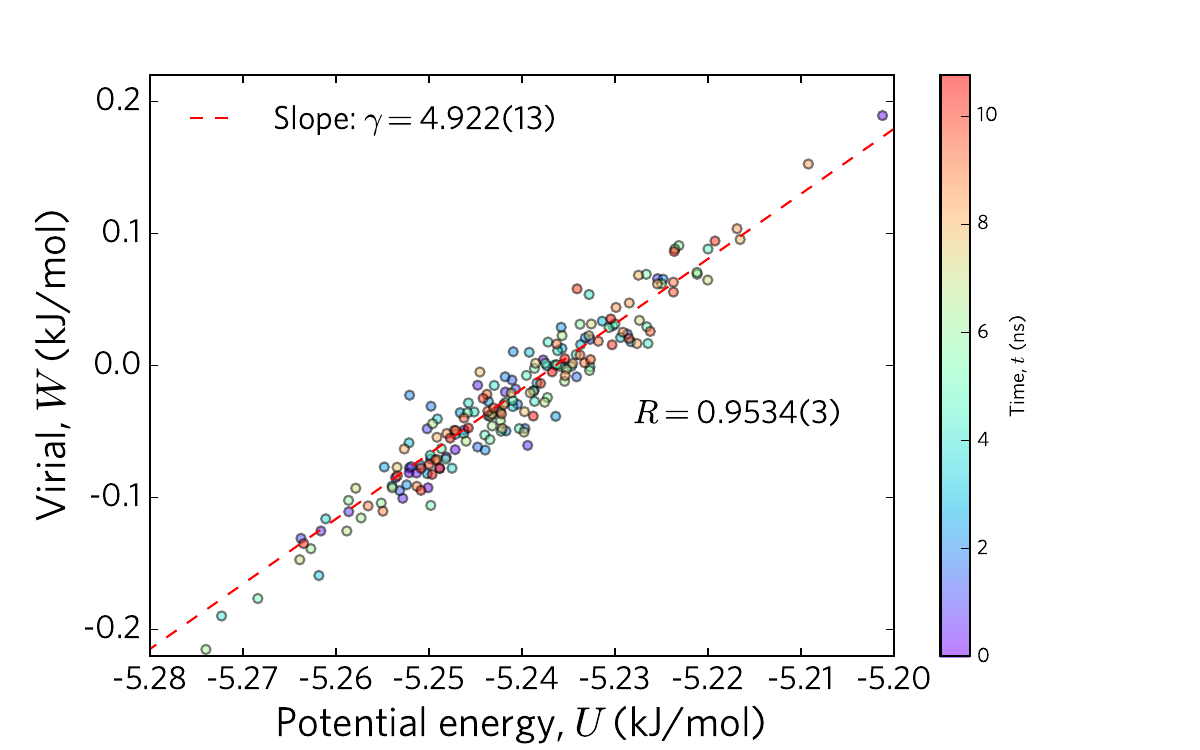}
\caption{\label{fig:lennard_jones_WU_scatter} Examination of instantaneous correlations between virial ($W$) and potential energy ($U$) fluctuations in the single-component Lennard-Jones model, as reported in Ref. \onlinecite{Pedersen2010b}. For clarity, 200 evenly spaced points are shown out of the $10^6$ total set. The $WU$ fluctuations are strongly correlated, with $R=0.9534(3)$ and slope $\gamma=4.922(13)$. }
\end{figure*}

Figure \ref{fig:lennard_jones_corr}(b) shows that the temporal CG scaling exponent $\gamma_{\bar U \bar W}$ estimated from the time-averaging approach yields a $\gamma$ value between 4.9 and 5.0 with statistical uncertainties of $2\%$. These $\gamma$ values are consistent with reported values from the literature that demonstrate scale invariance \cite{Pedersen2008}. To further gauge the performance of our CG correlation functions quantitatively, we estimated $R$ and $\gamma$ from the instantaneous correlation in Fig. \ref{fig:lennard_jones_WU_scatter}. The obtained correlation coefficients, $R=0.9534(3)$ and $\gamma=4.922(13)$, highlight that the temporal CG approach can accurately capture the hidden scale invariance of the Lennard-Jones potential energy surface. Overall, this additional analysis of the Lennard-Jones system further extends the demonstrated framework beyond simple repulsive interactions, thereby emphasizing the broader applicability of the proposed coarse-graining approach.

\bibliography{manuscript_v2}